\documentclass[]{spie}

\usepackage[]{graphicx}
\usepackage{amsmath, amssymb, amsfonts}
\usepackage[english]{babel}
\usepackage[]{epstopdf}
\usepackage{booktabs,caption,fixltx2e}
\usepackage[flushleft]{threeparttable}
\usepackage[font={small,it}, margin = 0.5cm]{caption}

\usepackage{subcaption}

\title{Active Correction of Aperture Discontinuities (ACAD) for space telescope pupils: a parametric analysis} 

\author{Johan Mazoyer\supit{a}, Laurent Pueyo\supit{a}, Colin Norman\supit{b}, Mamadou N'Diaye\supit{a}, Dimitri Mawet\supit{c}, R\'e{}mi Soummer\supit{a}, Marshall Perrin\supit{a}, \'E{}lodie Choquet\supit{a}, Alexis Carlotti\supit{d}
\skiplinehalf
\supit{a} Space telescope Science Institute, 3700 San Martin Drive, Baltimore, MD 21218, USA\\
\supit{b} Department of Physics and Astronomy, Johns Hopkins University, Baltimore, MD, USA\\
\supit{c} Astronomy Department, Caltech University, Pasadena, CA, USA\\
\supit{b} Institut de Plan\'e{}tologie et d'Astrophysique de Grenoble, France
}

\authorinfo{Further author information: contact Johan Mazoyer at jmazoyer@stsci.edu}
 
\begin{document} 
  
\maketitle 

\begin{abstract}
As the performance of coronagraphs improves, the achievable contrast is more and more dependent of the shape of the pupil. The future generation of space and ground based coronagraphic instruments will have to achieve high contrast levels on on-axis and/or segmented telescopes. To correct for the high amplitude aberrations introduced by secondary mirror structures and segmentation of the primary mirror, we explore a two deformable mirror (DM) method. The major difficulty of several DM methods is the non-linear relation linking actuator strokes to the point spread function in the coronagraph focal plane. The Active Compensation of Aperture Discontinuities (ACAD) method is achieving this minimization by solving a non linear differential Monge Ampere equation. Once this open loop method have reached the minimum, a close-loop stroke minimization method can be applied to correct for phase and amplitude aberrations to achieve the ultimate contrast. In this paper, I describe the results of the parametric analysis that that I have undertaken on this method. After recalling the principle of the method, I will described the explored parameter space (deformable mirror set-up, shape of the pupil, bandwidth, coronagraph designs). I will precisely described the way I simulated the Vortex coronagraph for this numerical simulation. Finally I will present the preliminary results of this parametric analysis for space telescope pupils only.
\end{abstract}



\section{Introduction}
\label{sec:Intro}

The coronagraph designs for the current generation of high contrast ground based instruments \cite{Beuzit08,Macintosh08,Hinkley11} were mostly designed for circular, unobstructed pupils. In some case \cite{Soummer11}{}, they may have taken into account the central obscuration. This was enough at the targeted level of contrast ($\sim 10^{-5}$), but the new frontier in the quest for the highest contrast levels is now the correction of the large diffractive artifacts effects introduced at the science camera by apertures of increasing complexity. Indeed, the future generation of space (WFIRST-AFTA\cite{Spergel2015}, ATLAST\cite{Postman2012,Feinberg2014} or HDST\cite{Delcanton2015}) and ground based coronagraphic instruments (EPICS@E-ELT\cite{Kasper10}{}, or the TMT\cite{Macintosh06}) will be mounted on on-axis and/or segmented telescopes: the design of coronagraphic instruments for such observatories is currently a domain undergoing rapid progress. 

Specific coronagraphic designs \cite{Carlotti14,Ndiaye15} are currently developed to reach high contrast levels after such pupils. These techniques will be used in addition with deformable mirrors (DMs) to correct for wavefront errors, either residuals of the adaptive optic system on ground based telescopes, or introduced by the optics themselves for space and ground based telescopes. The current designs for future high contrast instruments now systematically include two sequential DMs for the simultaneous correction of phase and amplitude wavefront errors.

However, several coronagraph \cite{Kuchner02,Guyon05,Mawet05,Soummer05} have already been developed for circular axi-symetric apertures, with various inner working angle and contrast performance. Another approach therefore consists of using these coronagraph ant the possibilities of the already provided sequential DMs to correct for aberrations introduced by secondary mirror structures and segmentation of the primary mirror.

In this paper, we present the Active Correction of Aperture Discontinuities (ACAD) technique, which was introduced by Pueyo \& Norman (2013)\cite{Pueyo_Norman13}{}. This method numerically resolve the Monge-Ampere second order differential equation which relate any given wanted apodization to the shapes of the DMs needed to produce it. In particular, it can be applied to find the DM shapes needed to correct for any given pupil discontinuities. Associated with a stroke minimization algorithm\cite{Pueyo09}{}, we can therefore converge to DM shapes producing very high contrast levels in the focal plane of the coronagraph. After a proof of concept on several pupils\cite{Pueyo14}{}, we are now aiming at a comprehensive analysis of the parameter space of this method. The contrast level is an important indicator of performance. However, we also want to analyze the effect of ACAD on other criteria: size of the bandwidth, inner and outer working angles (IWA and OWA), final throughput of the 2 DM + coronagraph system. We explore a large range of parameters, from the form of the pupil, the size and number of actuators of the DM and distance between them, to the type of coronagraph. We hope that a better understanding of the effects of these parameters on the performance of the system will help constrain the design of future high contrast instruments.

In Section~\ref{sec:method_descr}, we briefly recall the principle of the ACAD correction associated with the stroke minimization algorithm. In Section~\ref{sec:parameter_space}, we present the principal parameters of this study. We will described carefully the way we simulated the vortex coronagraph in this study in Section~\ref{sec:vortex_simulation}. Finally, we will present the preliminary results of the parametric study in Section~\ref{sec:results}.

\begin{figure}[ht]
 \begin{center}
  \includegraphics[width = 0.98\textwidth]{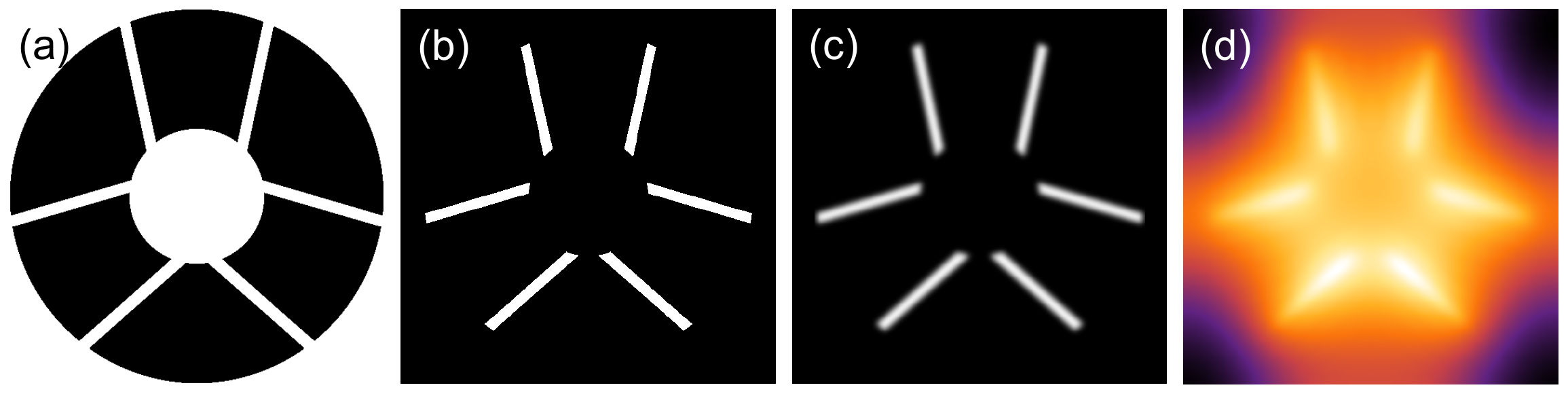}
\end{center}
 \caption[ACAD steps] 
{ \label{fig:acad_steps} Creating the ACAD apodization. (a) Itinitial pupil (WFIRST AFTA in this example). (b) Isolate the discountinuities of the pupil. (c) Tamper these discountinuities by a Gaussian filter. (d) Resulting DM shape after ACAD correction.}
\end{figure}

\section{A three step algorithm}
\label{sec:method_descr}

The solution for the DMs is obtained in three successive steps: calculation of the ACAD solution, light propagation through the coronagraph, and DM shape adjustment using the wavefront control algorithm.

We first calculate the best possible remapping solution, hereafter named the ACAD correction, by numerically solving the Monge-Ampere equation. The details of this solver are described in Pueyo \& Norman (2013)\cite{Pueyo_Norman13}{}. We multiply the initial pupil (for example the WFIRST AFTA pupil, see image (a) in Figure~\ref{fig:acad_steps}) by a binary mask to ensure that our algorithm  will concentrate its efforts on the support structures of the secondary and to prevent it from correcting the diffraction effects due to the central obstruction or the pupil edge (leaving such operations to the coronagraph and its apodization). The result is presented on the image (b) in Figure~\ref{fig:acad_steps}. To reduce the stroke applied on the actuators, we taper the discontinuities that we want to correct in the pupil by a Gaussian filter (image (c) in Figure~\ref{fig:acad_steps}). An example of the resulting DM shape is presented in the image (d) of Figure~\ref{fig:acad_steps}. 

This is an ``open loop'' solution and our solver uses a pupil plane estimate and not a post-coronagraph image plane based metric. For this reason, the algorithm is not coronagraphic dependent. The Monge Ampere equation is also resolve independent of the set up of the DMs (diameter, distance between them) or event the wavelength. The contrast improvement during this phase is limited to a factor 10 to 100. \\

In a second step, we propagate the solution of the two DM surfaces through the APLC coronagraph using the Fresnel approximation to obtain the contrast in the final image plane. We proved in \cite{Mazoyer_JATIS} that this propagation is more accurate than the SR-Fresnel propagation in this case. We often reduce the size of the pupil relatively to the DM, which appears on the image (b) (c) and (d) compared to (a), in Figure~\ref{fig:acad_steps}. This reduces the actual number of actuators in the pupil, leaving a part of the mirror outside of the pupil. However, this is often necessary as the imaged pupil on the second DM plane is larger than the actual pupil. If we take a pupil of the same size than the DMs, the propagated electrical field in the second DM plane might end up larger than the DM itself. This will create a discontinuity in the field that is going to create a "cross" artifact in the final focal plane. Although this effect have the characteristic shape of a discontinuity in a Fourier Transformed plane, note that it is not a numerical effect. The influence of the DM set up on this effect will be studied on Section~\ref{sec:comparsetup}.\\

Finally, we use a close loop quasi-linear algorithm called Stroke minimization (Pueyo et al., 2009 \cite{Pueyo09}{}) to adjust the DM surfaces and obtain the final contrast. Starting from the ACAD solution, this algorithm seeks a two-DM solution that improves the contrast in the image plane by 5\% at each iteration, while minimizing the stroke of the actuators. This algorithm is particularly useful in this specific case, as important strokes have already been put on the DM actuators with the ACAD solution to correct for the discontinuities in the pupil. In this third step, we assume a perfect estimation of the complex electric field in the focal plane for both the monochromatic and 10\% broadband light cases. Several methods have been developed to retrieve an estimate of this field\cite{Borde06,Giveon07,Baudoz06,Mazoyer14}{}, which are not specific to the 2 DM correction.\\

Due to the limited number of actuators in the pupil, we are only allowed to correct for a small part of the focal plane, called the dark hole (DH). In addition, we limit the correction to an annulus of dimensions ranging between an inner working angle (IWA) and a outer working angle (OWA). The number of iterations is variable, as the algorithm stops when the correction starts to diverge. In the current implementation of our algorithms, we apply the same weight to every actuator in this correction. Because some of these actuators are hidden behind the structures of the secondary, this might not be the optimal solution.

\section{Explored parameter space}
\label{sec:parameter_space}

In this section, we describe the parameters used in our study. We tested our algorithm on 4 different space telescope pupils: the ATLAST pupil, the WFIRST-AFTA pupil, a WFIRST-AFTA like pupil with half-width spiders, and an off-axis segmented pupil. In each case, we made a correction in monochromatic light and in 10\% bandwidth. We do not introduce phase or amplitude errors. Indeed, several algorithms have been developed to correct for these aberrations. The goal of this method is to find DM shapes for which the pupil discontinuity effects in the focal plane are negligible compared to other effects. 

We put a special attention to the DM setup and the description of the 2 coronagraphs in the next sections.

\subsection{Deformable mirrors}
\label{sec:DMsetup}

\begin{table}[]
\centering
\caption{Deformable Mirror setup}
\label{tab:DMsetup}
\begin{tabular}{l c c c}
                            & HiCAT case & AFTA like case & Intermediate case  \\   \hline  \hline                       
\# of actuators              & 34         & 34             & 34               \\ 
DM Inter-act. distance & 0.3 mm        & 1 mm             & 0.3 mm              \\ 
Size DM D            & 1 cm         & 3.5 cm          & 1 cm                \\ 
Inter DM distance Z      & 0.3 m        & 1 m              & 1 m                 \\ 
$D^2/Z$                   & 0.3 mm        & 0.1 mm            & 1 mm                \\ \hline
\end{tabular}
\end{table}

The ACAD solution general shape does not depend on the DM setup (size D of the DMs, distance Z between them) or on the wavelength. However, the strokes are linearly related to the ratio $D^2/Z$ which make it a very important parameter of our study. We describe in this section the 3 different setups we analyzed covering a large $D^2/Z$ range.

First the number of actuators is fixed to 34. This corresponds to the number of actuators on the Boston Micromachines DMs used on the High Contrast Imaging (HiCAT) Bench (see companion paper by N'Diaye et al. in these proceedings for a description of the HiCAT bench). We study the two cases of a "Boston Micromachines" like DM (inter-actuator distance of ~0.3 mm) and of a "Xinetics" like DM (inter-actuator distance of ~1 mm). This inter-actuator distance and the given number of actuators constrain the size of the DM to D =1 cm for the Boston Micromachines like DM and D = 3.5 cm for the Xinetics like DM. We then studied 2 inter DM distances for the Boston Micromachines DM (Z = 1 m and Z = 0.3 m) and one inter DM distance for the Boston Micromachines DM (Z = 1 m and Z = 0.3 m). 
The first setup (Boston Micromachines like DM, D = 1 cm, and Z = 0.3 m) is the setup of the HiCAT bench and we will refer to it as HiCAT case. The second setup (Xinetics like DM, D = 3.5 cm, and Z = 1 m) is the chosen setup for the WFIRST-AFTA bench and we will refer to it as AFTA case. We also test an intermediate case (Boston Micromachines like DM, D = 1 cm, and Z = 1 m). The parameters of these 3 different setups, covering a large $D^2/Z$ range, are reported in Table~\ref{tab:DMsetup}.

As mentioned in the previous section, we reduce the diameter of the pupil compared to the DM size. Therefore, out of the initial 34 actuators, we only use 30 actuators across the pupil diameter. This will reduce the size of the DH, limiting the OWA to $15\lambda/D$.

\subsection{Coronagraphy}
\label{sec:corono}

We simulated two coronagraphs in this study: an apodized pupil Lyot coronagraph (hereafter APLC, Soummer et al. 2005\cite{Soummer05}, N'Diaye et al. 2015\cite{Ndiaye15}) and a ring apodized vortex coronagraph (hereafter vortex, Mawet et al. 2005\cite{Mawet05}, Mawet et al, 2015\cite{Mawet13}). 

The characteristics of these two coronagraphs are reported in Table~\ref{tab:corono}. The APLC coronagraph use an optimized apodization associated to a classical Lyot mask to produce a $10^{-9}$ contrast level in a for $r<40 \lambda/D$ over a 10\% bandwidth for a pupil with a central obstruction (but with no spiders). The IWA is determined by the size of the Lyot mask (of radius $5 \lambda/D$ for the AFTA pupil and $3.75 \lambda/D$ for the ATLAST pupil). The vortex is an achromatic coronagraph. The ring apodization is optimized to obtain a analytically null contrast in the DH. We describe how we simulate this coronagraph in the next section.

\begin{table}[ht]
\centering
\begin{threeparttable}
\caption{Vortex and APLC coronagraph parameters}
\label{tab:corono}
\begin{tabular}{lcc}
                                 & AFTA case                      & ATLAST case               \\
                           \hline      \hline     
Pupil COR$^a$        & 0.36 R                                                                                                                      & 0.2 R                                                                                                                         \\ \hline 
Vortex ring apodization          & \begin{tabular}[c]{@{}c@{}}t=1 in 0.36 R\textless r \textless 0.71 R\\ t=0.67 in 0.67 R\textless r \textless R\end{tabular} & \begin{tabular}[c]{@{}c@{}}t=1 in 0.2 R\textless r \textless 0.51 R\\ t= 0.85 in 0.51 R\textless r \textless R\end{tabular} \\
APLC focal mask radius           & 5 $\lambda$/D       & 3.75 $\lambda$/D                         \\
Vortex Lyot COR$^a$ & 0.67 R            & 0.4 R                \\
APLC Lyot COR$^a$  & 0.5 R      & 0.51 R                \\
\hline
\end{tabular}
\begin{tablenotes}
      \small
      \item[$^a$] Central obscuration radius (compared to the outer radius R of the pupil)
    \end{tablenotes}
\end{threeparttable}
\end{table}

\section{Simulating a vortex coronagraph}
\label{sec:vortex_simulation}
\begin{figure}[ht]
 \begin{center}
   \includegraphics[height = 0.3\textwidth, trim= 0cm 0cm 0cm 0cm, clip = true]{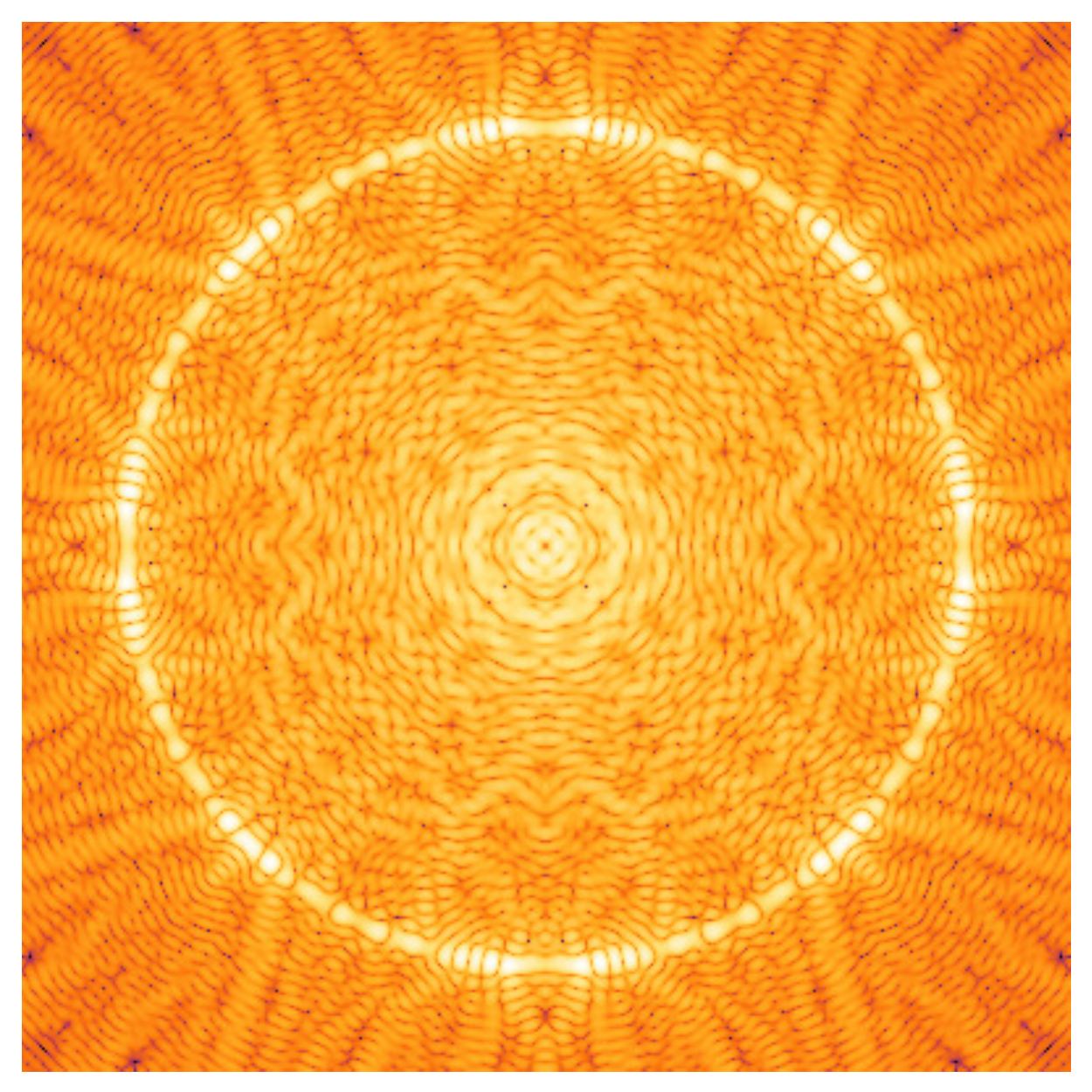}
  \includegraphics[height = 0.3\textwidth, trim= 0cm 0cm 0cm 0cm, clip = true]{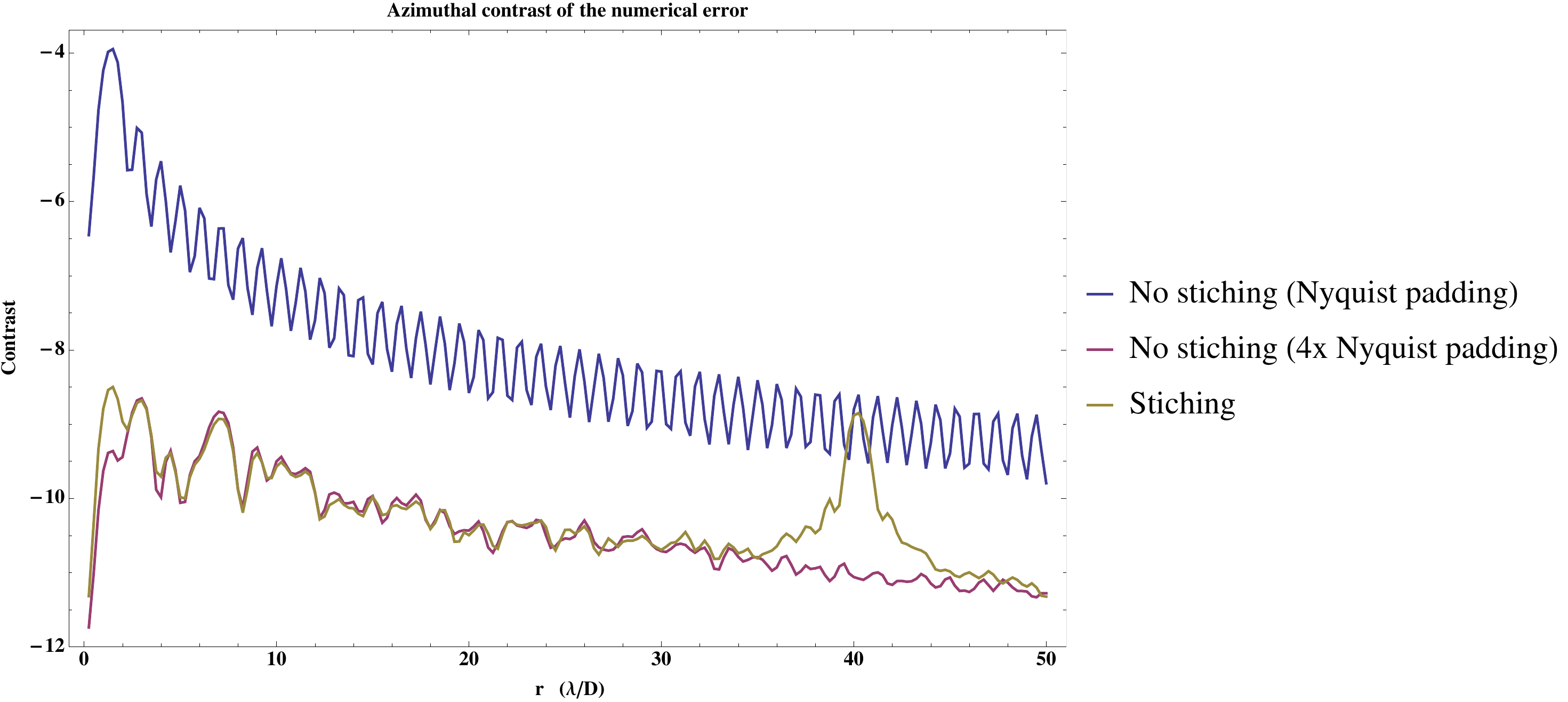}
\end{center}
 \caption[] 
{ \label{fig:num_error_vortex_stich} Left : Image of the intensity after a coronagraph numerically simulated with ``stichings''. The field producing this intensity is analytically null and must be removed during the numerical simulation of the vortex. Right: Radial profile of the azimuthal contrast obtained with a vortex coronagraph where the mask is: a N$\lambda$/D vortex mask with a Nyquist padding (in blue), a N$\lambda$/D vortex with a 4 times Nyquist padding (in red), a N$\lambda$/D vortex with a 20 times Nyquist padding between 0 and 80$\lambda$/D and a Nyquist padding between 80 and N $\lambda$/D (in yellow).}
\end{figure}

We present in this section the method we used to simulate the full focal plane of a vortex coronagraph for complex apertures combining the Matrix Fourier Transform (MFT, see Soummer et al. 2007\cite{Soummer07}{}) algorithm with different samplings depending on the frequency.

We first take the case of a second order coronagraph with a circular aperture. In the absence of aberrations, the coronagraph perfectly reject the light outside of the Lyot pupil and the resulting field in the focal plane is analytically null. However, due to numerical errors, this field is not null. We therefore have to remove this intensity in the final Lyot plane to obtain a null field in the vortex focal plane. However, the resulting field of an aberration in the pupil will be convoluted by this error in the final focal plane, and we need to find a solution to minimize this error before subtracting it.

The sampling of the mask to be at the Nyquist limit must be at least of 2 pixels per $\lambda/D$. It results that to access all frequencies, we have, for a pupil of N pixels, we have at the minimum to over-pad the size of the focal plane by a factor of two (size of the focal plane in pixel: 2xN). However, the intensity in the focal plane is represented by the blue curve in Figure~\ref{fig:num_error_vortex_stich} (right). We see that even with this solution is Nyquist compatible, we are left with a very high numerical error. One simple solution is to increase the sampling in the focal plane. The red curve in Figure~\ref{fig:num_error_vortex_stich} (right) is obtained by using a sampling of 8 pixels per $\lambda/D$ (size of the focal plane in pixel: 8xN). We see that the resulting numerical error is drastically reduce. However, this solution is very CPU demanding, especially if N is important. 

The solution we propose, is to separate the field incoming on the vortex mask in two parts: the inner part (0 to 40 $\lambda/D$) will have a sampling of 20 pixel per $\lambda/D$ (size of the focal plane 80*20 pixels). For this case, the MFT algorithm is very effective. The rest of the frequencies will have a Nyquist sampling (size of the focal plane 2*N pixels). The 2 fields are then summed in the Lyot plane. This operation is only slightly more CPU demanding than the Nyquist limit solution. The resulting intensity is represented in Figure~\ref{fig:num_error_vortex_stich} (left). The impact of this "stitching" in focal plane is obvious. However, we show that for most of the frequencies the resulting intensity (in yellow in Figure~\ref{fig:num_error_vortex_stich}, right) is a lot inferior to the Nyquist sampling case, for a CPU time only slightly superior. At the "stitching" frequency ($40 \lambda/D$) the numerical error is important. However, we managed to locate this feature far outside the DH possible zone to avoid any problem during the correction. We adopt this method to quickly simulate the full focal plane field after a Vortex coronagraph. 

To simulate the second order ring apodized vortex coronagraph, the same method is applied, only with the fact that the "analytically null field" in this case is the one of an un-aberrated wavefront through a pupil with a central obstruction and ring apodization and a Lyot stop also with a central obstruction (see Mawet et al. 2013\cite{Mawet13}{} for details).

\section{Preliminary results of the parametric analysis}
\label{sec:results}

\begin{figure}[]
 \begin{center}
 \begin{subfigure}[a]{0.98\textwidth}
        \includegraphics[width = 0.47\textwidth, trim= 0cm 4.5cm 5cm 4cm, clip = true]{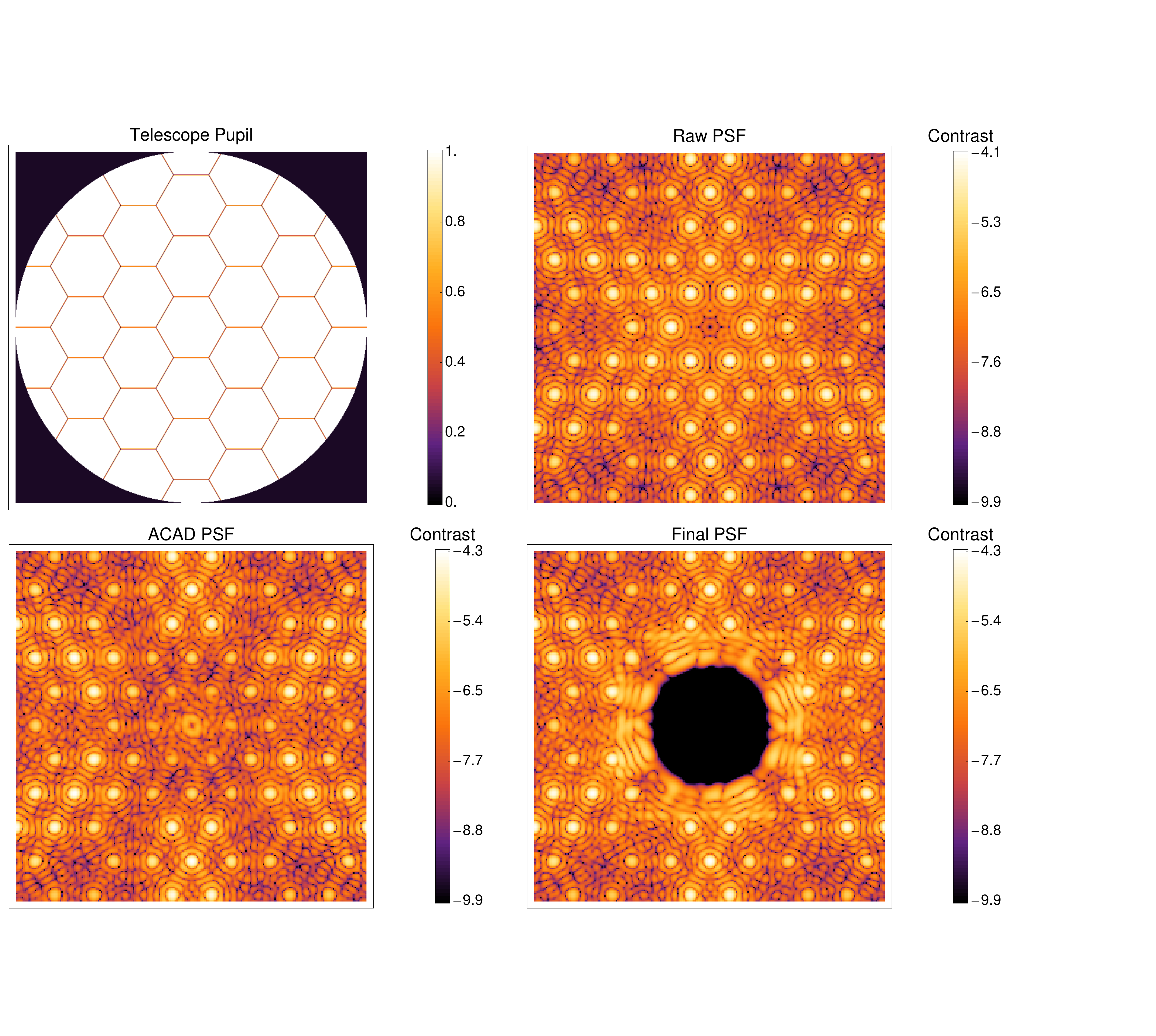}
        \includegraphics[width = 0.47\textwidth, trim= 0cm 3cm 3cm 3cm, clip = true]{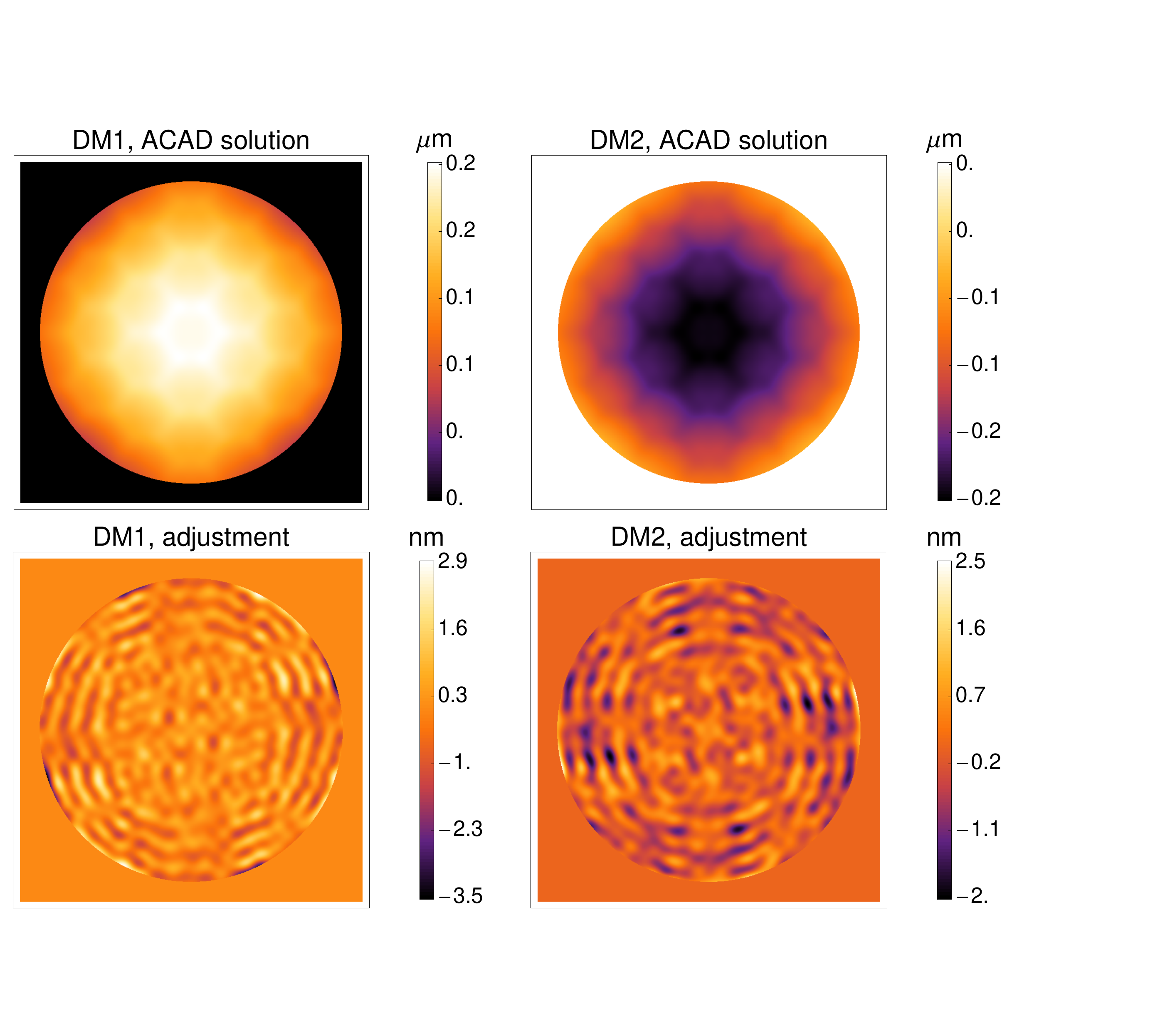}
    \caption{Off axis segmented pupil and a vortex coronagraph}
    \label{fig:focalpupilplane_OAS}
\end{subfigure}

 \begin{subfigure}[b]{0.98\textwidth}
        \includegraphics[width = 0.47\textwidth, trim= 0cm 4.5cm 5cm 4cm, clip = true]{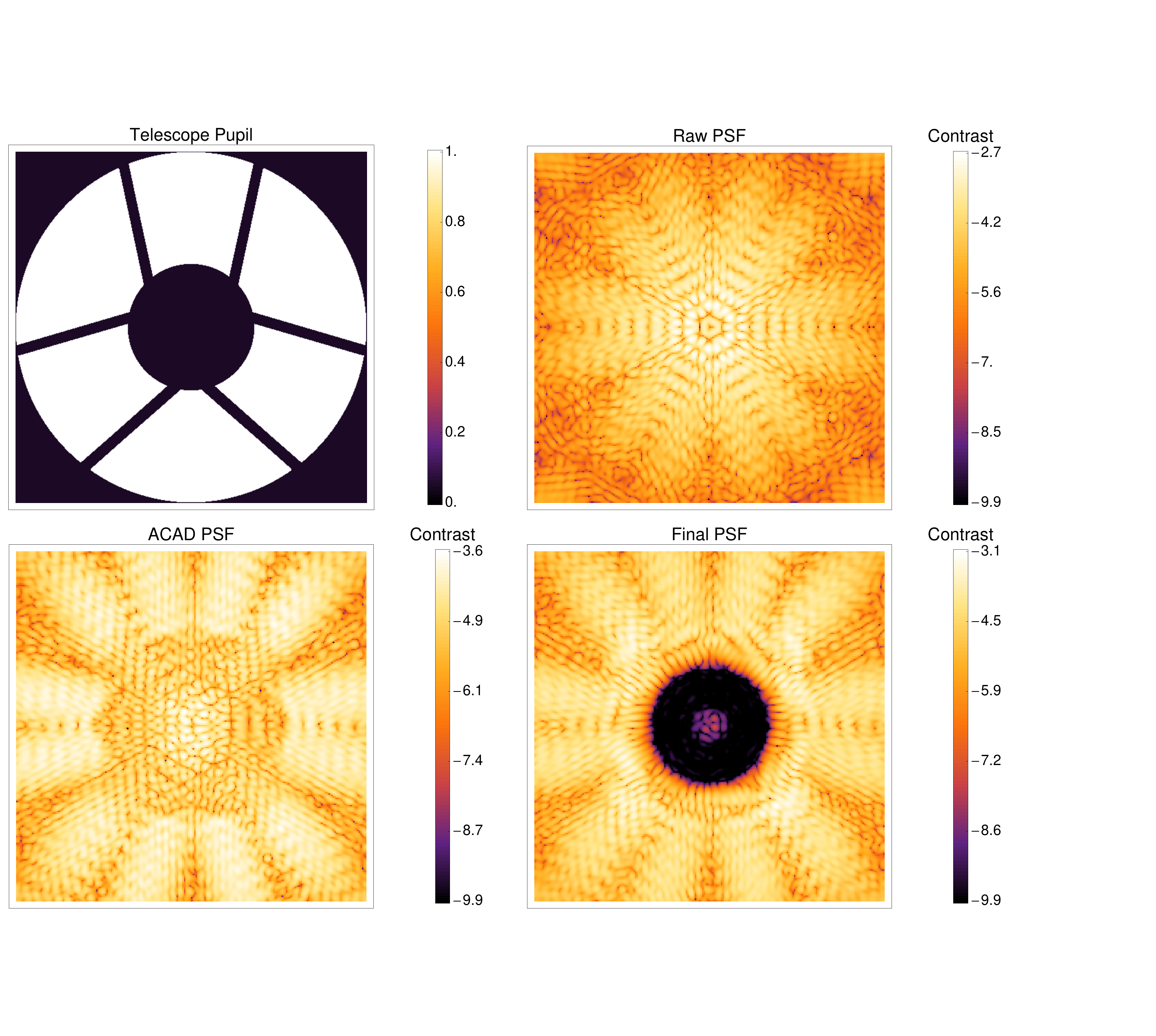}
        \includegraphics[width = 0.47\textwidth, trim= 0cm 3cm 3cm 3cm, clip = true]{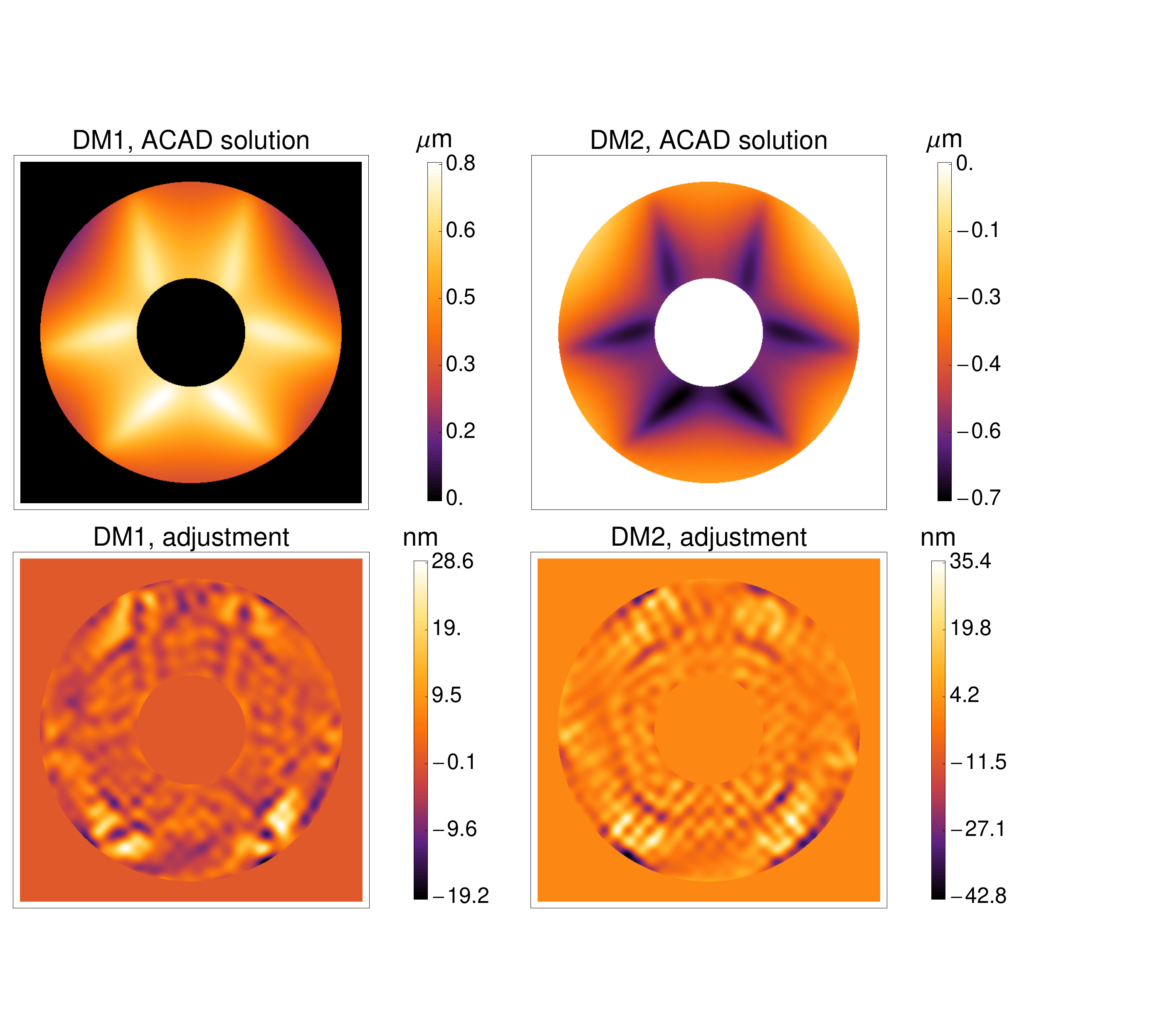}
    \caption{WFIRST-AFTA pupil and a vortex coronagraph}
    \label{fig:focalpupilplane_AFTA}
\end{subfigure}

 \begin{subfigure}[c]{0.98\textwidth}
        \includegraphics[width = 0.47\textwidth, trim= 0cm 4.5cm 5cm 4cm, clip = true]{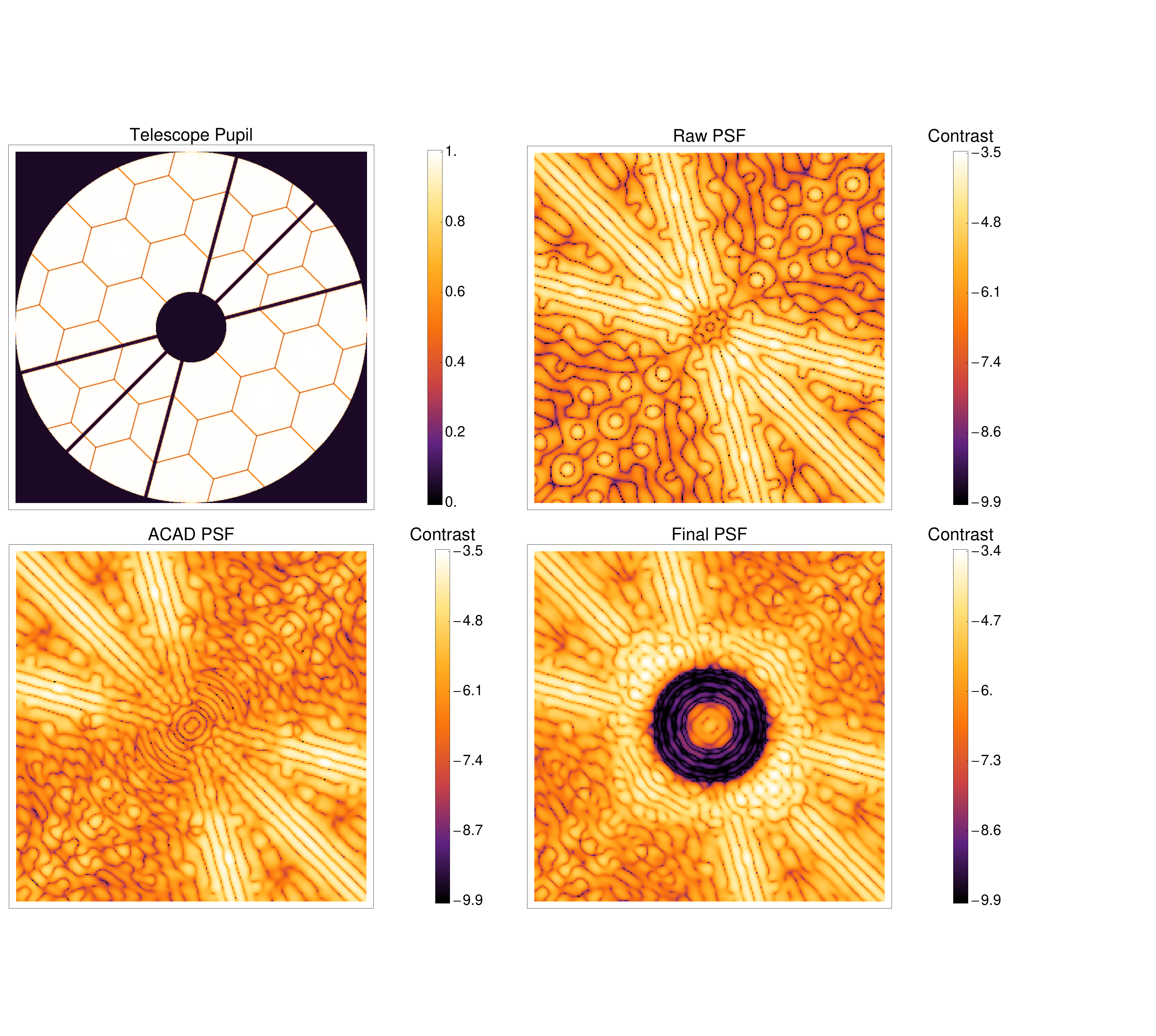}
        \includegraphics[width = 0.47\textwidth, trim= 0cm 3cm 3cm 3cm, clip = true]{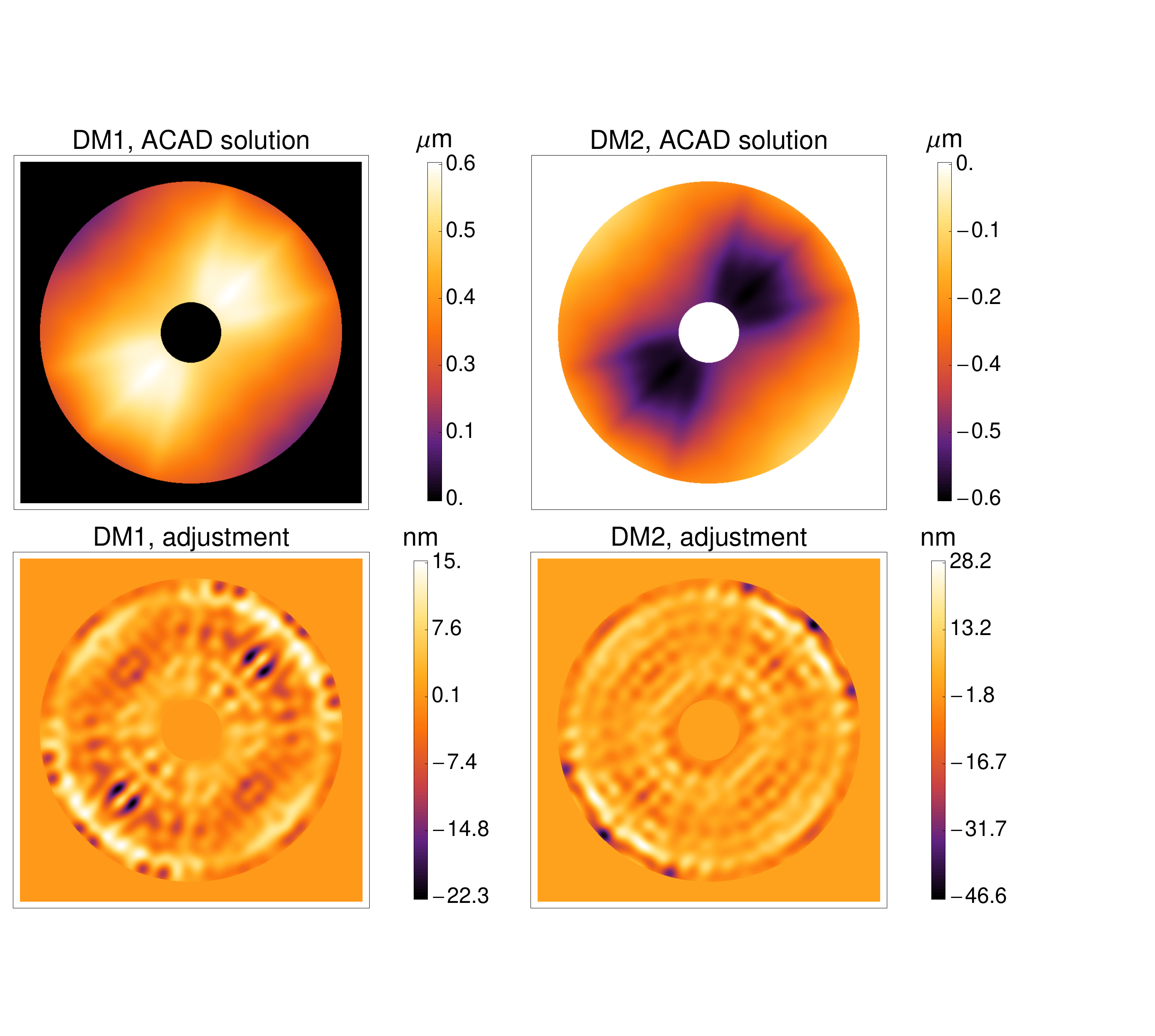}
    \caption{ATLAST pupil and an APLC}
    \label{fig:focalpupilplane_ATLAST}
\end{subfigure}
\end{center}
 \caption[Focal plane after the differe] 
{ \label{fig:focalpupilplanes} 
For each pupils. Left: Initial pupil and focal planes at different steps of the correction for this pupil. 
Right: DM shapes and strokes after the ACAD step (top) and after the adjustments by the stroke minimization method (bottom).}
\end{figure}

In this section, I present the preliminary results that I obtained for both coronagraph designs for the different pupils. Unless stated otherwise, all the simulation were carried out using the "HiCAT bench" case : D = 1cm, Z = 0.3m.

\subsection{Off-axis segmented pupil}
\label{sec:offaxis_results}

\begin{figure}[ht]
 \begin{center}
  \includegraphics[width = 0.48\textwidth, trim= 1.3cm 0cm 1.3cm 0cm, clip = true]{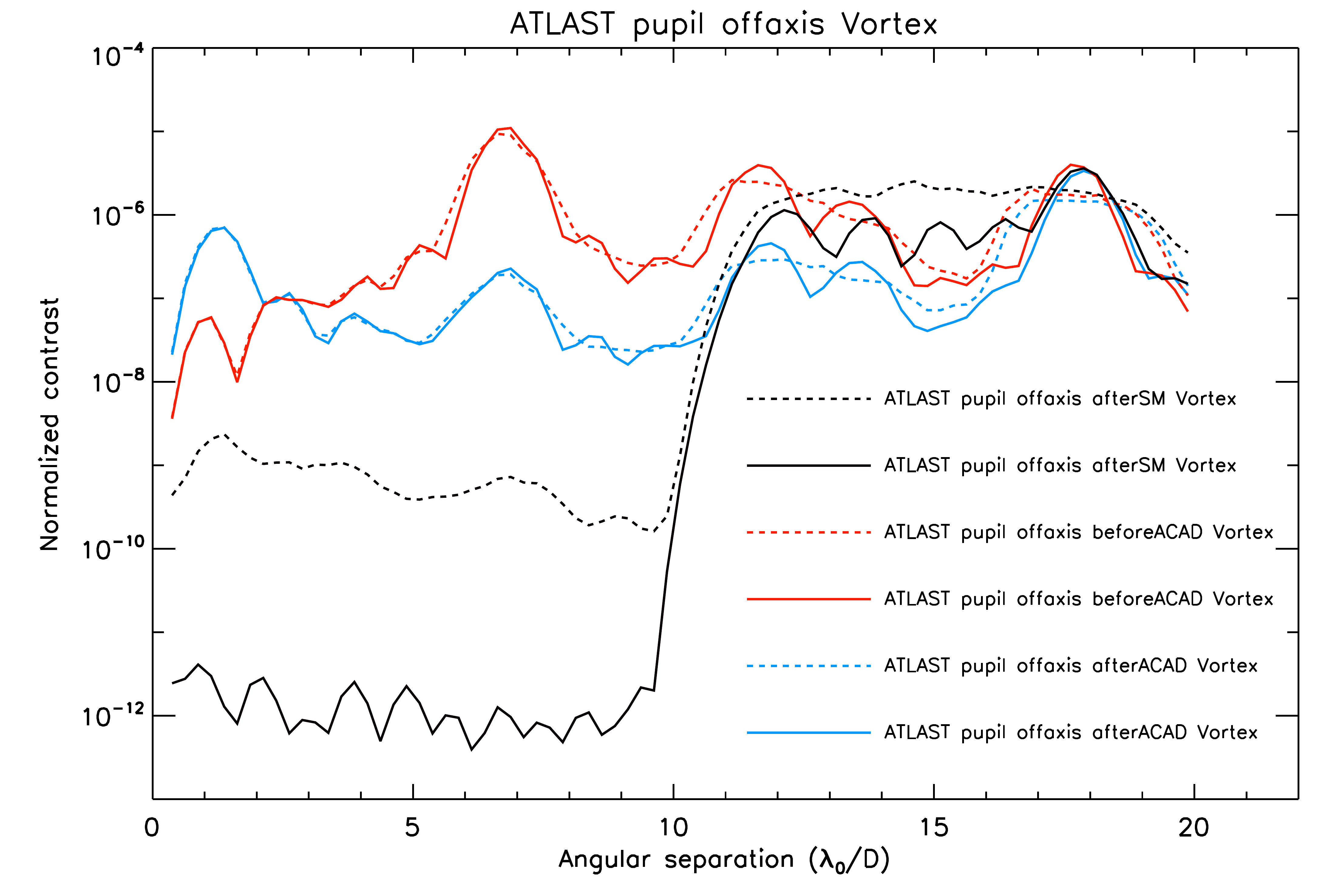}
\end{center}
 \caption[] 
{ \label{fig:contrastcurve_offaxis_Vortex} Radial profile of azimuthally averaged contrast curves for the correction on the OAS pupil as a function of the distance to the star in $\lambda/D$. We use a Vortex coronagraph in a 1 to 10 $\lambda/D$ DH. The DMs are 1 cm large and situated at a distance of 0.3 m. The solid curves are in monochromatic light, the dashed curves are in 10\% bandwidth light. In red, we plot the contrast before any correction. In blue, we plot the contrast after the ACAD correction. Finally, in black, we plot the final contrast after ACAD + Stroke minimization.}
\end{figure}

We simulate the ACAD correction an off-axis segmented (OAS) pupil, followed by a stroke minimization over a 1 to 10 $\lambda/D$, in monochromatic and polychromatic light. We only used the non apodized vortex coronagraph in this case. Indeed, in the absence of central obscuration, it is not useful to use ring apodization nor APLC designs. Figure~\ref{fig:focalpupilplane_OAS} (left) shows the OAS pupil (top left) and the focal plane after each step of the correction. On top right, we see the initial focal plane after the vortex coronagraph, in the case of two flat DMs. We then show the focal plane after the ACAD solution (bottom left). The contrast have not improved a lot, but we see that within the frequencies achievable by the DMs, we corrected for the diffraction peaks created by the segmentation. Finally, we create a very deep DH in this region after the stroke minimization. The corresponding DM shapes are represented in Figure~\ref{fig:focalpupilplane_OAS} (right). On top we represent the shapes given by the ACAD solution for DM1 and DM2. On bottom, we represent the adjustments to this shape made by the stroke minimization algorithm. In the case of the OAS pupil, these adjustments are very small ($<3$ nm).

The corresponding radial profile of azimuthally averaged contrast curves are reported in Figure~\ref{fig:contrastcurve_offaxis_Vortex}. The solid curves are in monochromatic light, the dashed curves are in 10\% bandwidth light. In red we represent the contrast before any correction (flat DMs), in blue the contrast after ACAD, in black the contrast after ACAD + stroke minimization. We obtained results of $5.10^{-12}$ in monochromatic light and of $5.10^{-10}$ in polychromatic light for the whole DH (0 to 10 $\lambda/D$).

\subsection{WFIRST-AFTA pupil}
\label{sec:AFTA_results}

\begin{figure}[ht]
 \begin{center}
  \includegraphics[width = 0.48\textwidth, trim= 1.3cm 0cm 1.3cm 0cm, clip = true]{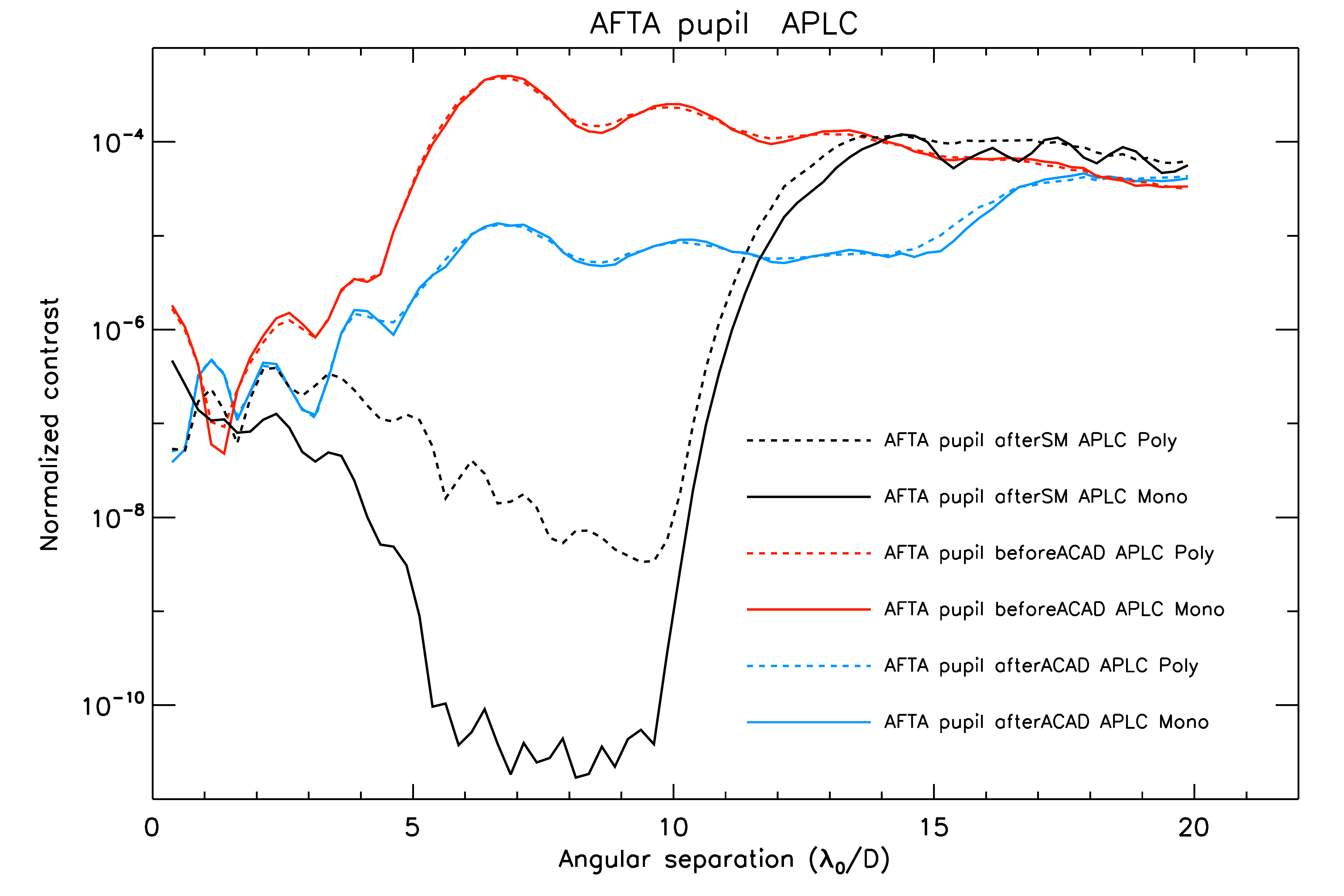}
  \includegraphics[width = 0.48\textwidth, trim= 1.3cm 0cm 1.3cm 0cm, clip = true]{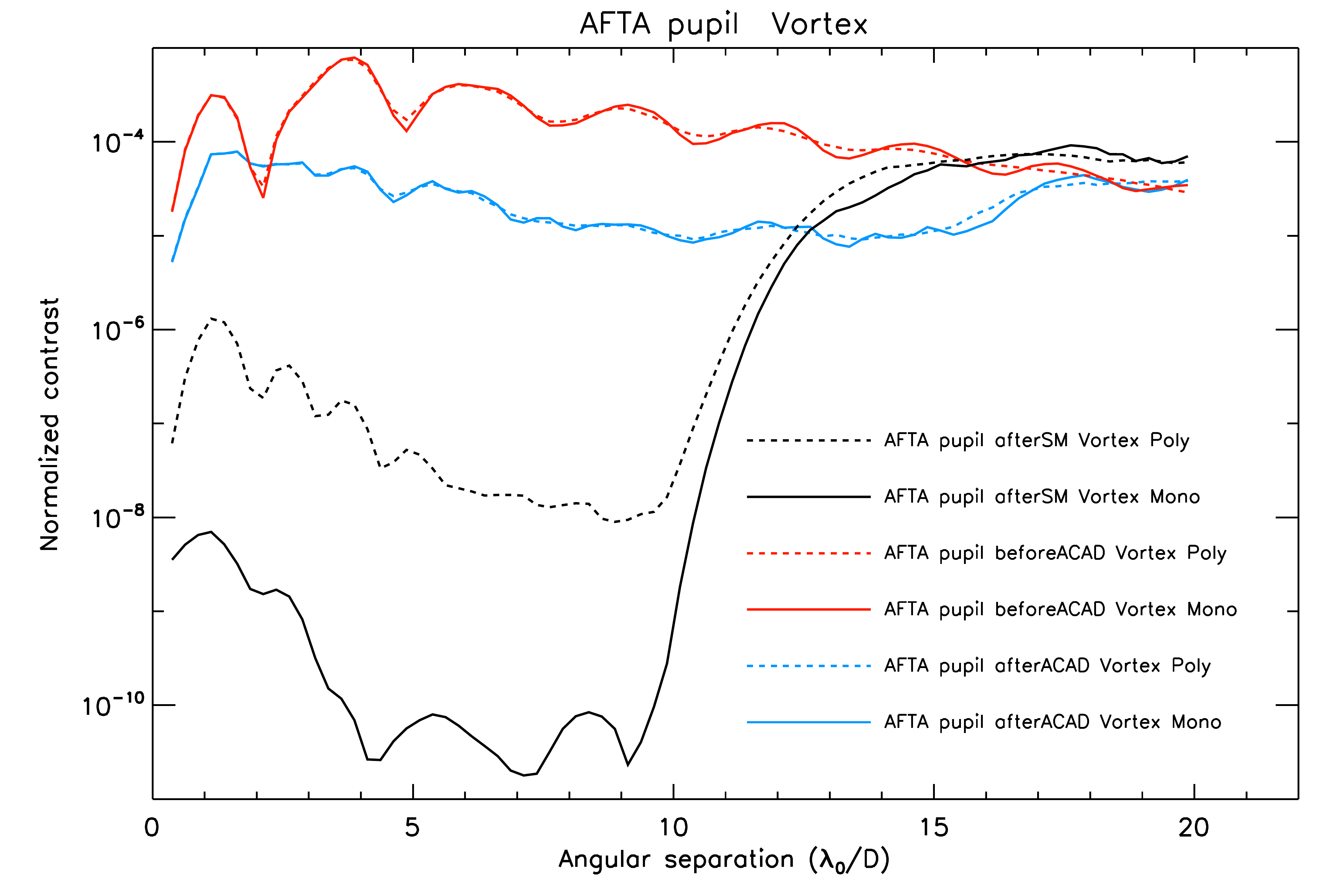}
\end{center}
 \caption[] 
{ \label{fig:contrastcurve_AFTA_APLC_Vortex} Radial profile of azimuthally averaged contrast curves for the correction on the AFTA pupil as a function of the distance to the star in $\lambda/D$. In the left plot, we use a APLC coronagraph in a 5 to 10 $\lambda/D$ DH. In the right plot, we use a Vortex coronagraph in a 3 to 10 $\lambda/D$ DH. The DMs are 1 cm large and situated at a distance of 0.3 m. The solid curves are in monochromatic light, the dashed curves are in 10\% bandwidth light. In red, we plot the contrast levels before any correction. In blue, we plot the contrast levels after the ACAD correction. Finally, in black, we plot the final contrast after ACAD + stroke minimization.}
\end{figure}

In this section, we present the result obtained with a WFIRST-AFTA like pupil. Like in the previous section, we first present the pupil that we used (Top left, Figure~\ref{fig:focalpupilplane_AFTA}) and the focal plane of the vortex coronagraph at each step of the correction (Figure~\ref{fig:focalpupilplane_AFTA}) for a vortex coronagraph. We then show the DM shapes after the ACAD correction (Figure~\ref{fig:focalpupilplane_AFTA}, right top) and the adjustments made after the stroke minimization algorithm (Figure~\ref{fig:focalpupilplane_AFTA}, right bottom). Finally, we show the corresponding radial profile of azimuthally averaged contrast curves in Figure~\ref{fig:contrastcurve_AFTA_APLC_Vortex}. With an APLC, on a 5-10 $\lambda/D$ DH (Figure~\ref{fig:contrastcurve_AFTA_APLC_Vortex}, left) we obtained a $8.10^{-11}$ contrast level in monochromatic light and of $1.10^{-8}$ in polychromatic light for the whole DH. With an Vortex (Figure~\ref{fig:contrastcurve_AFTA_APLC_Vortex}, right) we obtained a $9.10^{-11}$ contrast level in monochromatic light on a 3-10 $\lambda/D$ DH and of $2.10^{-8}$ in polychromatic light on a 4-10 $\lambda/D$ DH.

\begin{figure}[ht]
 \begin{center}
  \includegraphics[width = 0.48\textwidth, trim= 0cm 0cm 0cm 0cm, clip = true]{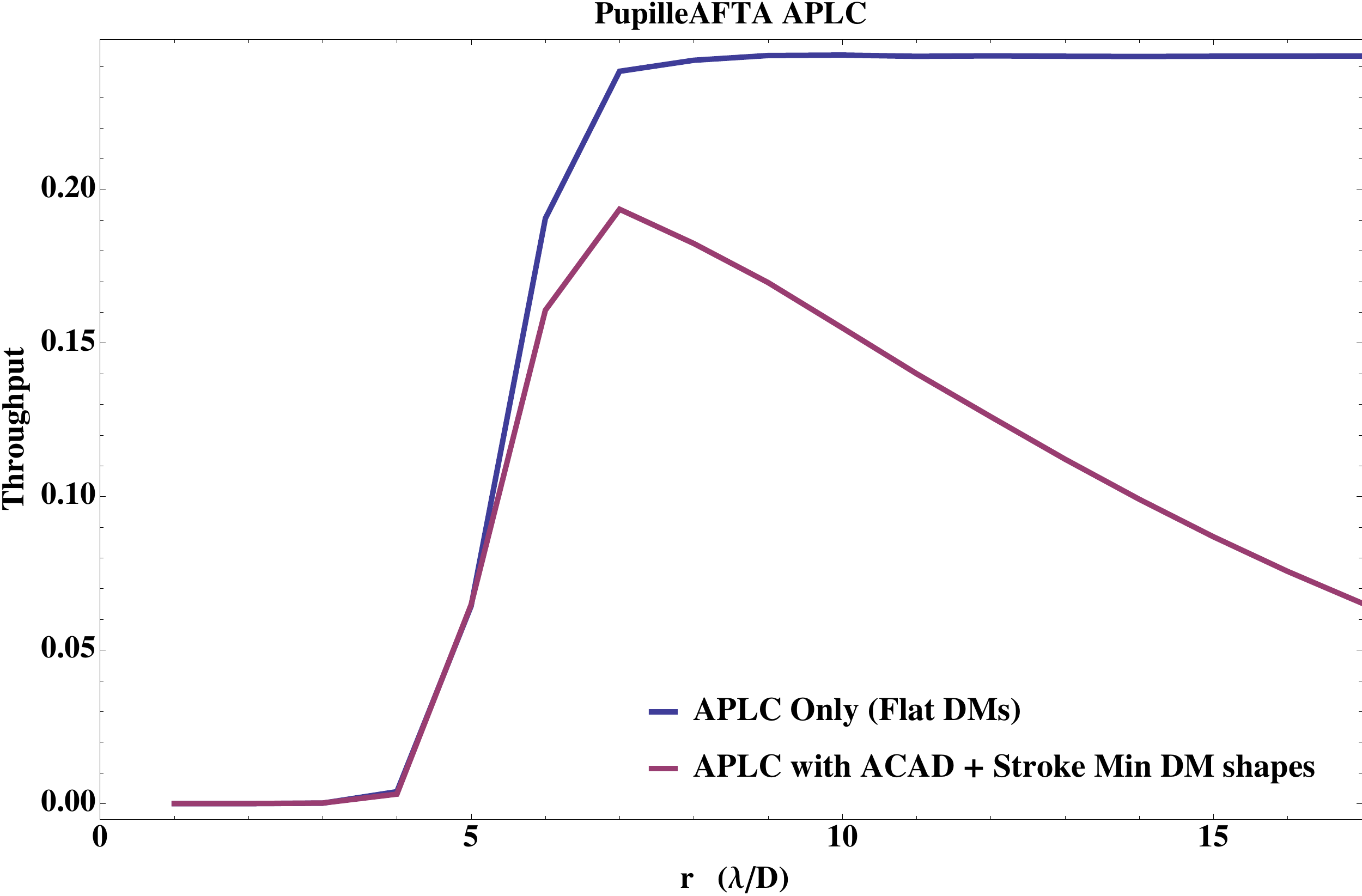}
  \includegraphics[width = 0.48\textwidth, trim= 0cm 0cm 0cm 0cm, clip = true]{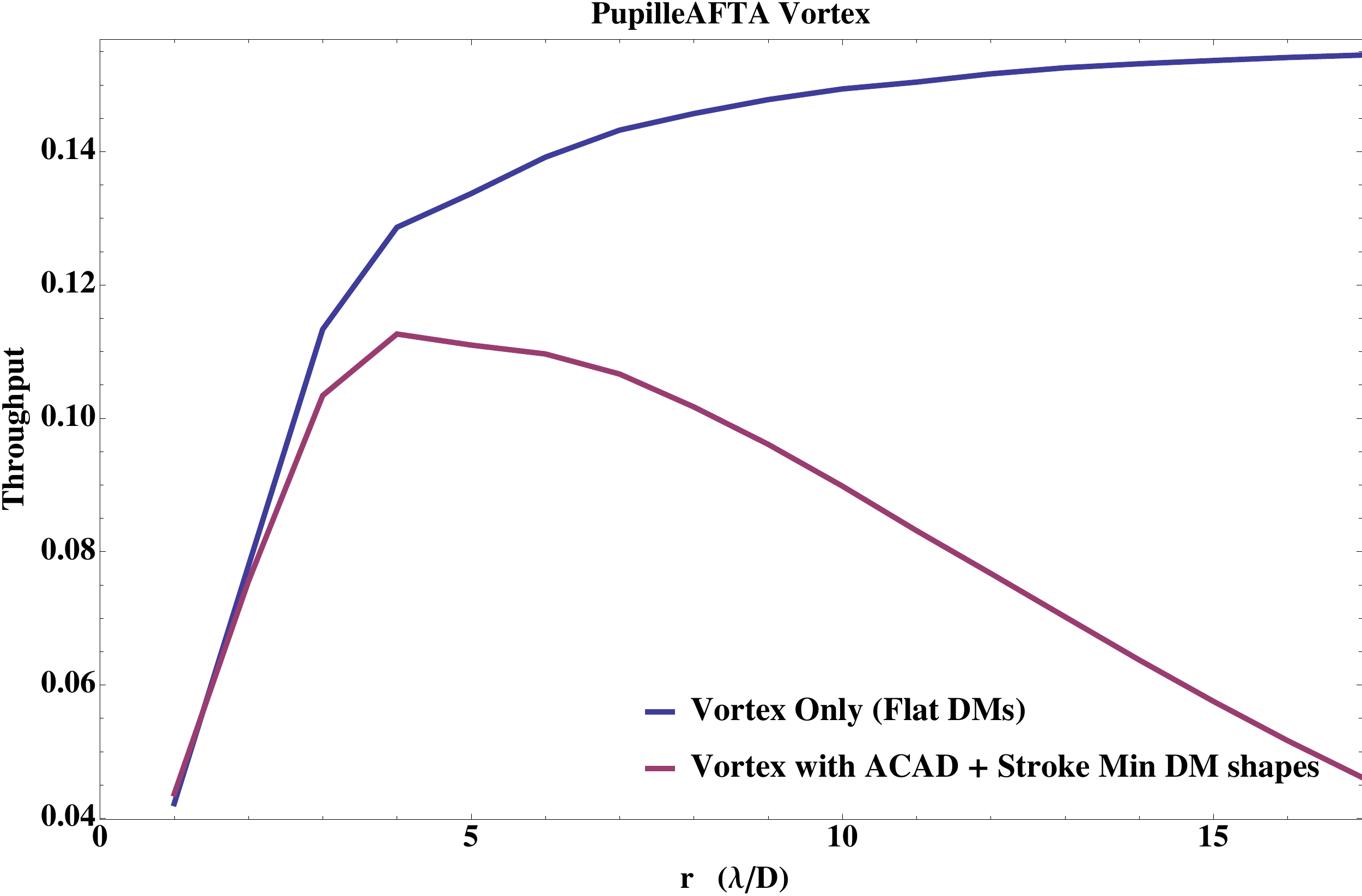}
\end{center}
 \caption[] 
{ \label{fig:throughput_AFTA_APLC_Vortex} Throughput curves for the correction on the AFTA pupil as a function of the distance to the star in $\lambda/D$, in the case of the APLC coronagraph (left) and Vortex coronagraph (right).}
\end{figure}

We present the throughput of our system in Figure~\ref{fig:throughput_AFTA_APLC_Vortex} at every radius of the star in $\lambda/D$. In blue, we plotted the throughput of the coronagraph alone (with the AFTA pupil and flat DMs). In red, we plot the final throughput when the shapes of the DMs shapes are set. The final throughput is almost only driven by the ACAD form. Therefore, it does not depend on the size and form of the DH.

For the APLC, we obtain a throughput better than 9\% on 5 to 15 $\lambda/D$. For the ring apodized Vortex coronagraph (right), we obtain a throughput better than 9\% 3 to 10 $\lambda/D$. The throughput for the Vortex coronagraph is particularly low (considering that this is a phase-only coronagraph). Indeed, as shown in Table~\ref{tab:corono}, the size of the Lyot central obstruction reaches 0.67 times the radius of the pupil for the AFTA pupil. 

The fact that the throughput is particularly low for high frequencies is due to the very large spiders of the WFIRST-AFTA pupil. Large spiders tend to introduce higher frequencies on the DMs that will degrade throughput at these frequencies, compared to the coronagraph alone. We did a correction with an WFIRST-AFTA like pupil with half-width spiders, which only slightly improved the results in contrast but results in a good improvement of the throughput at large frequencies.

\subsection{ATLAST}
\label{sec:ATLAST_results}

\begin{figure}[ht]
 \begin{center}
  \includegraphics[width = 0.48\textwidth, trim= 1.3cm 0cm 1.3cm 0cm, clip = true]{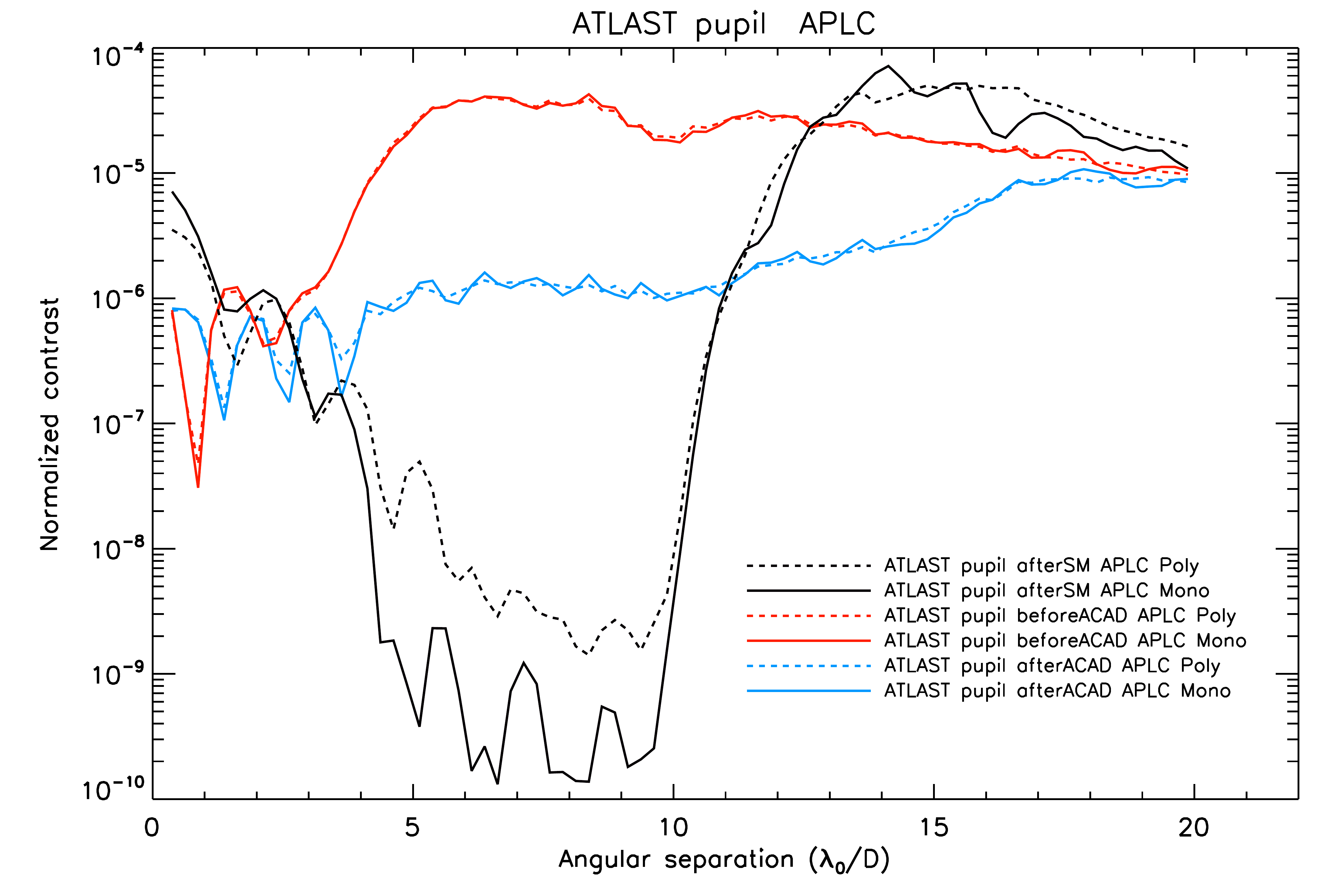}
  \includegraphics[width = 0.48\textwidth, trim= 1.3cm 0cm 1.3cm 0cm, clip = true]{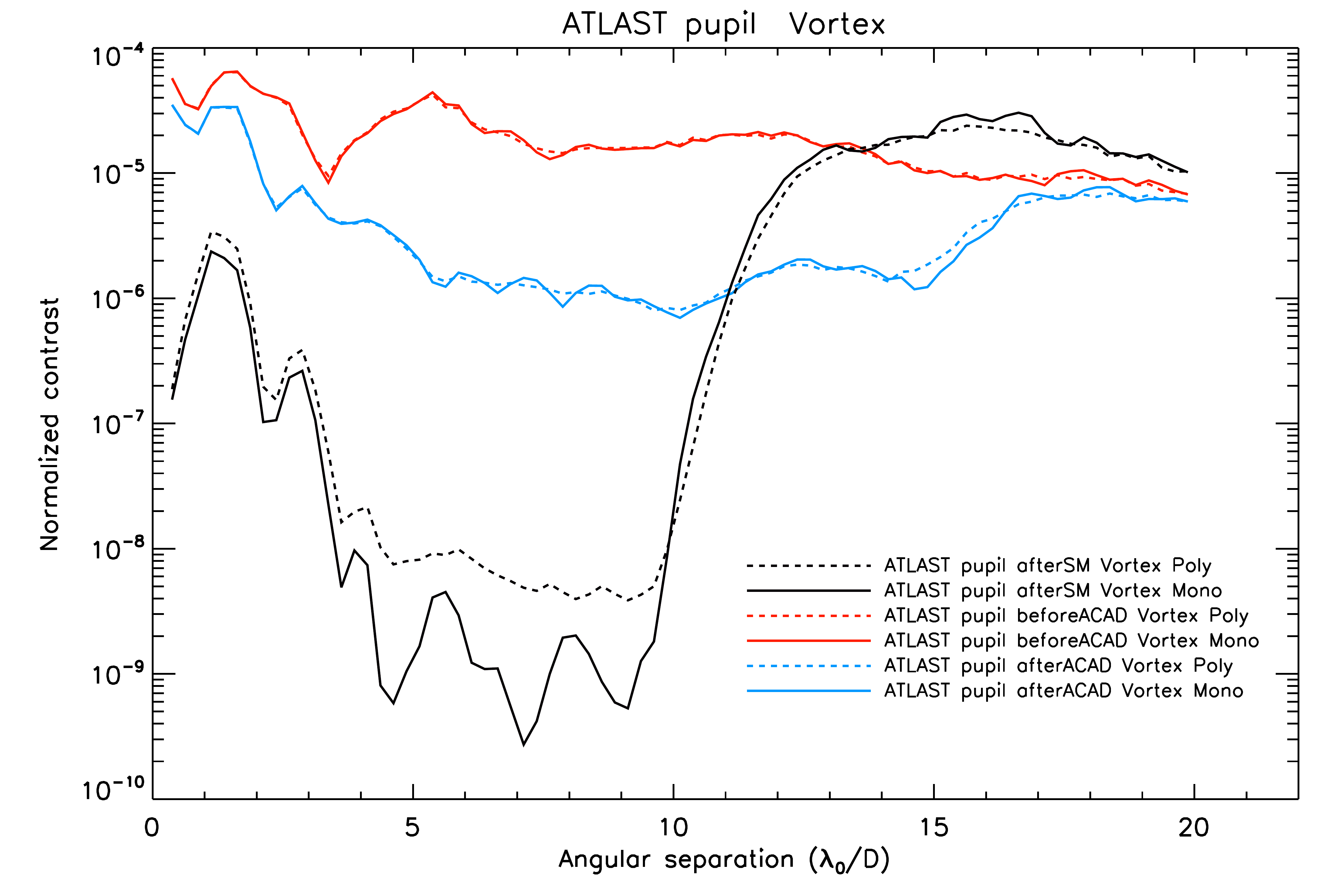}
\end{center}
 \caption[] 
{ \label{fig:contrastcurve_ATLAST_APLC_Vortex} Radial profile of azimuthally averaged contrast curves for the correction on the ATLAST pupil as a function of the distance to the star in $\lambda/D$. In the left plot, we use a APLC coronagraph in a 4 to 10 $\lambda/D$ DH. In the right plot, we use a Vortex coronagraph in a 3 to 10 $\lambda/D$ DH. The DMs are 1 cm large and situated at a distance of 0.3 m. The solid curves are in monochromatic light, the dashed curves are in 10\% bandwidth light. In red, we plot the contrast before any correction. In blue, we plot the contrast after the ACAD correction. Finally, in black, we plot the final contrast after ACAD + Stroke minimization.}
\end{figure}

In this section, we present the result obtained with a ATLAST like pupil. We first present the pupil that we used (top left, Figure~\ref{fig:focalpupilplane_ATLAST}) and the focal plane of the vortex coronagraph at each step of the correction (Figure~\ref{fig:focalpupilplane_ATLAST}) for an APLC coronagraph this time. We then show the DM shapes after the ACAD correction (Figure~\ref{fig:focalpupilplane_ATLAST}, right top) and the adjustments made after the Stroke minimization algorithm (Figure~\ref{fig:focalpupilplane_ATLAST}, right bottom), in the APLC case. Finally, we show the corresponding radial profile of azimuthally averaged contrast curves in Figure~\ref{fig:contrastcurve_ATLAST_APLC_Vortex}. With an APLC, on a 4-10 $\lambda/D$ DH (Figure~\ref{fig:contrastcurve_ATLAST_APLC_Vortex}, left) we obtained a $9.10^{-10}$ contrast level in monochromatic light and of $9.10^{-9}$ in polychromatic light for the whole DH. With an Vortex, on a 3-10 $\lambda/D$ DH (Figure~\ref{fig:contrastcurve_ATLAST_APLC_Vortex}, left) we obtained a $4.10^{-9}$ contrast level in monochromatic light and of $1.10^{-8}$ in polychromatic light for the whole DH.

\begin{figure}[ht]
 \begin{center}
  \includegraphics[width = 0.48\textwidth, trim= 0cm 0cm 0cm 0cm, clip = true]{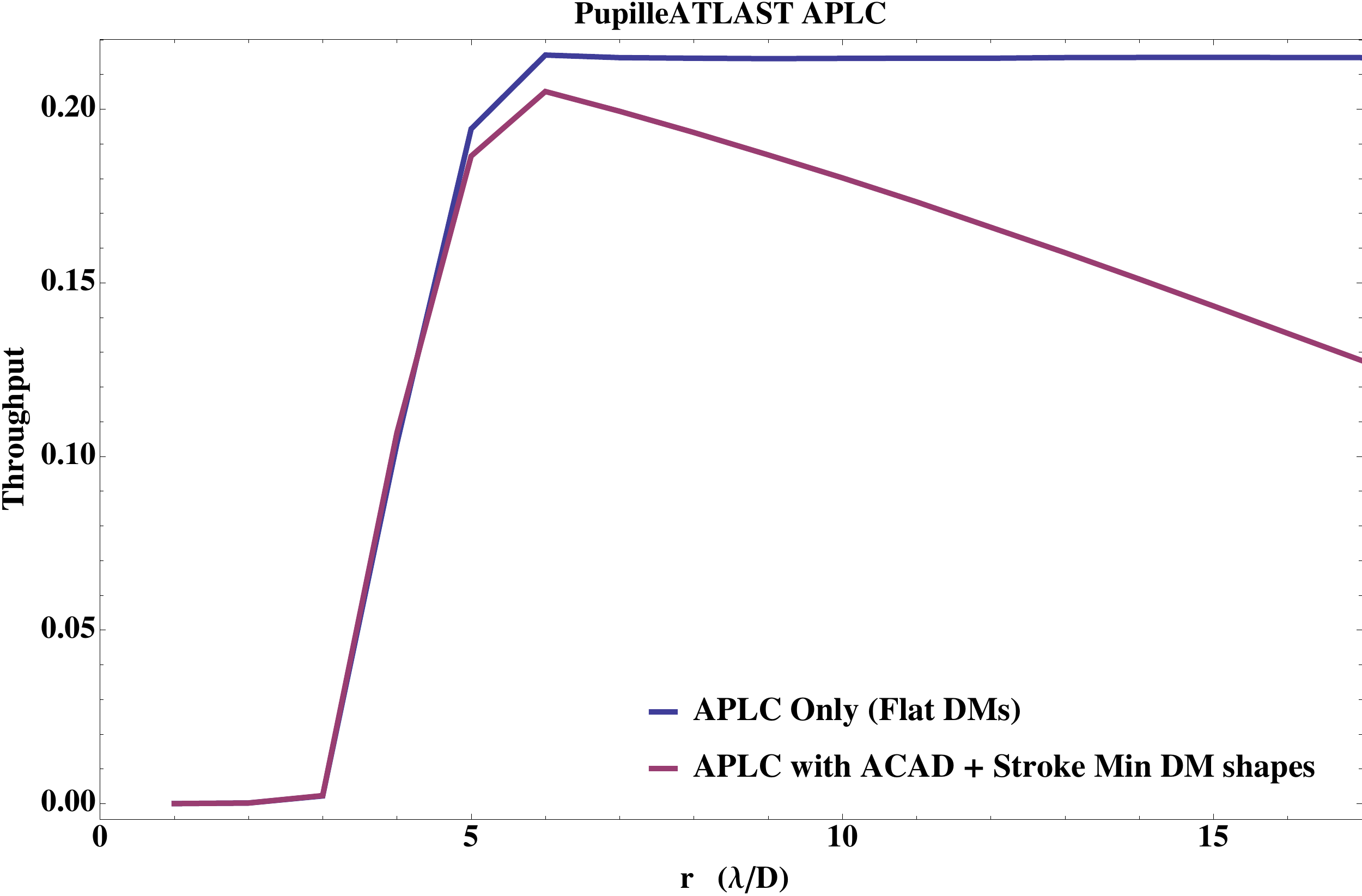}
  \includegraphics[width = 0.48\textwidth, trim= 0cm 0cm 0cm 0cm, clip = true]{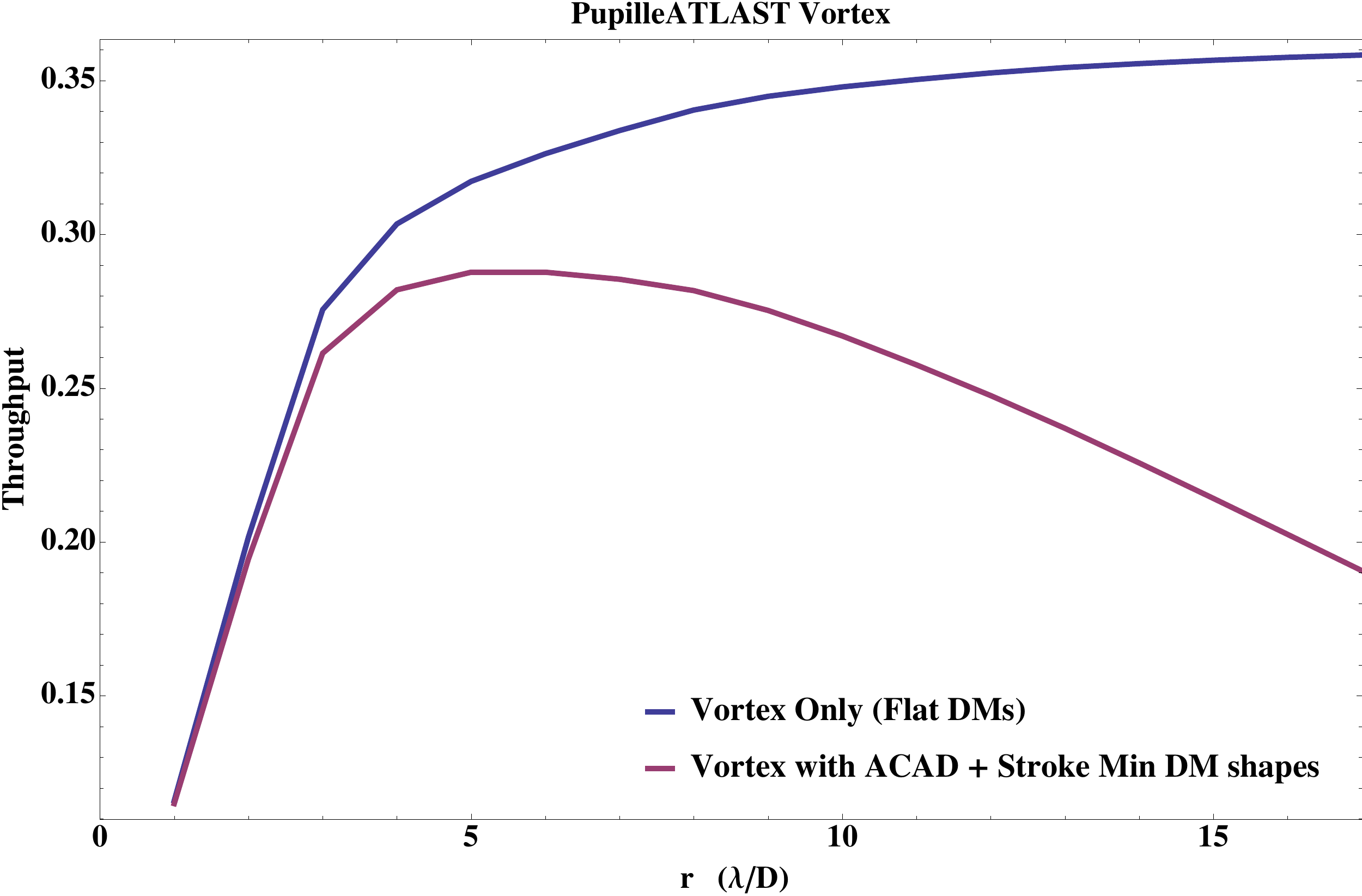}
\end{center}
 \caption[Contrast curve for an AFTA pupil with an APLC over a large DH] 
{ \label{fig:throughput_ATLAST_APLC_Vortex} Throughput curve for the correction on the ATLAST pupil as a function of the distance to the star in $\lambda/D$, in the case of the APLC coronagraph (left) and Vortex coronagraph (right).}
\end{figure}

We present the throughput of our system in Figure~\ref{fig:throughput_ATLAST_APLC_Vortex} at every radius of the star in $\lambda/D$. In blue, we plotted the throughput of the coronagraph alone (with the ATLAST pupil and flat DMs). In red, we plot the final throughput when the shapes of the ACAD DMs shapes are set. For the APLC, we obtain a throughput better than 13\% on 4.5 to 15 $\lambda/D$. For the ring apodized Vortex coronagraph (right), we obtain a throughput better than 21\% 3 to 15 $\lambda/D$.

\subsection{Current limitations}
\label{sec:limitations}

\begin{figure}[ht]
 \begin{center}
  \includegraphics[width = 0.48\textwidth, trim= 1.3cm 0cm 1.3cm 0cm, clip = true]{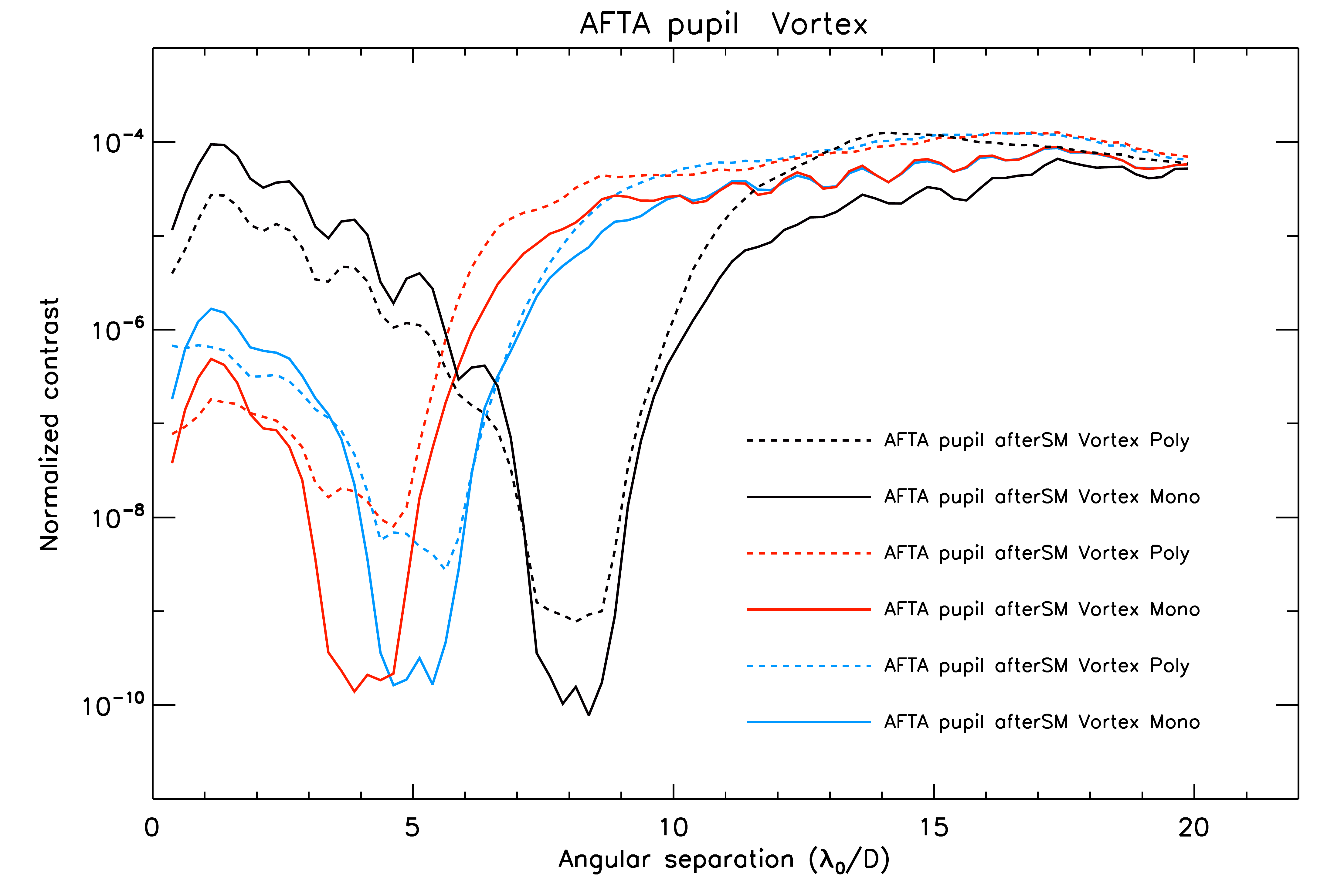}
  \includegraphics[width = 0.48\textwidth, trim= 1.3cm 0cm 1.3cm 0cm, clip = true]{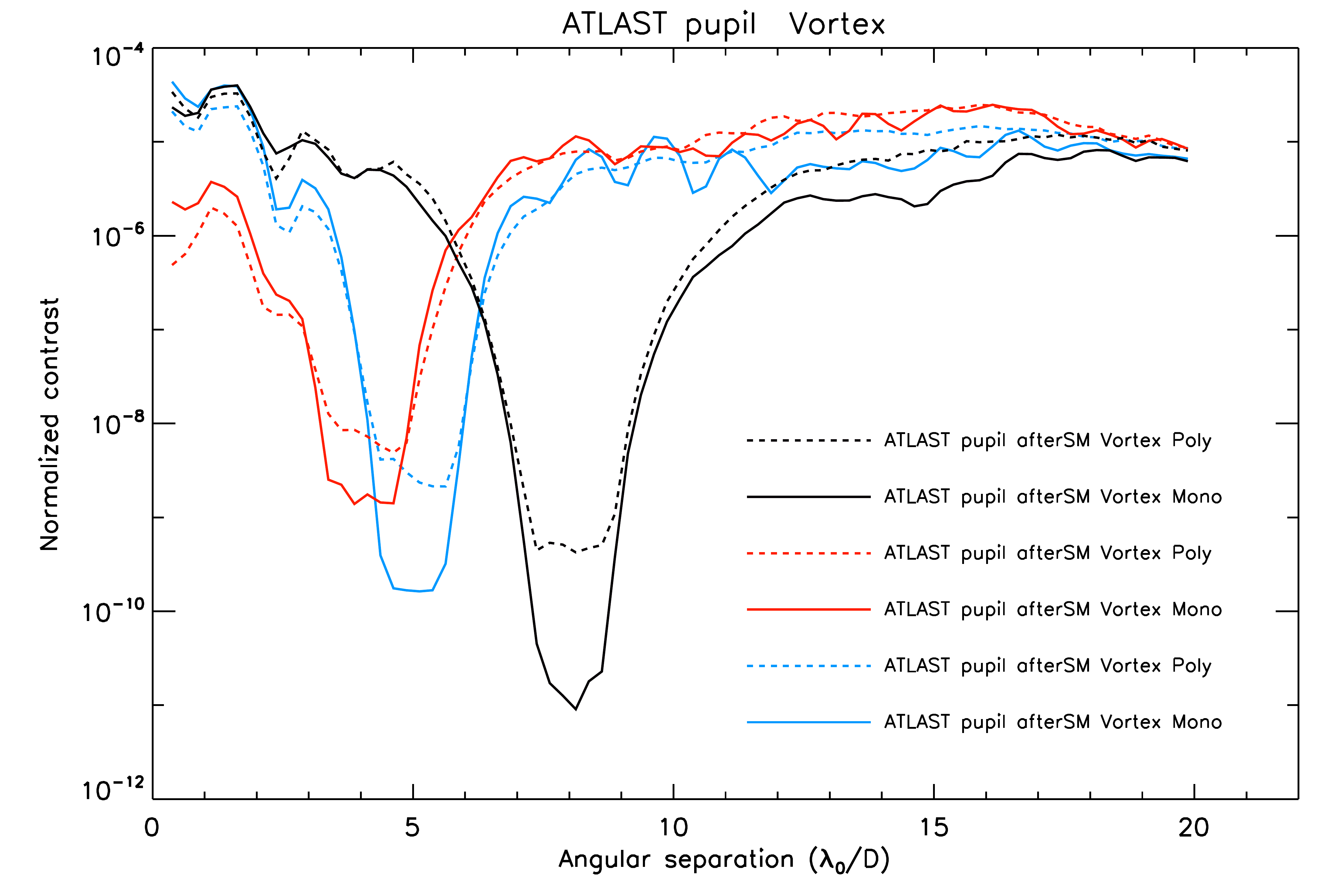}
\end{center}
 \caption[] 
{ \label{fig:contrastcurve_Vortex_smallDH} Contrast curve for the correction with a Vortex as a function of the distance to the star in $\lambda/D$ for the AFTA pupil (left) and ATLAST pupil (right).  The correction was conducted in 2 $\lambda/D$ DHs centered on 4, 5 and 8 $\lambda/D$. We represented only the final contrast after ACAD + Stroke minimization. The DMs are 1 cm large and situated at a distance of 0.3 m. The solid curves are in monochromatic light, the dashed curves are in 10\% bandwidth light.}
\end{figure}

First, we identified a limit in the number of actuators in the stroke minimization algorithm. Starting with the same ACAD shapes than in Sections~\ref{sec:AFTA_results}~and~\ref{sec:ATLAST_results}, we are able to improve the contrast locally in reducing the size of the DH to 2 $\lambda/D$. The results are presented in Figure~\ref{fig:contrastcurve_Vortex_smallDH} for the WFIRST-AFTA pupil (left) and ATLAST pupil (right), for the vortex coronagraph. The improvement of contrast when using a smaller DH, especially in polychromatic light, suggests that we can expect an improvement on a large DH when using more actuators in the pupil. We also expect to improve the contrast in polychromatic light by improving the stroke minimization correction.

The ACAD algorithm presented in Pueyo \& Norman (2013)\cite{Pueyo_Norman13} to resolve the Monge-Ampère equation ensures that the resulting apodization is the closest possible to the wanted apodization (image (c) in Figure~\ref{fig:acad_steps}) in the limits of the DM capabilities. However, there is maybe room for improvement of the algorithm by a more carefully defined "wanted apodization", as we are currently putting constrains on the wanted apodization outside the pupil and inside the central apodization. By doing so, we ensure that the algorithm is not trying to correct for the obstructed pupil, because this is normally done by the coronagraph itself (ring apodized vortex or APLC). Constraints on the apodization in these locations of the pupil are normally not necessary, because they will be blocked during the passage of the pupil, before the coronagraph. However, these out-of-pupil constrains have an influence of the strokes and on the level of "useful apodization" (inside the pupil) that we can obtained for a given DM with a finite number of degrees of freedom. 
We recently notice that changing the value of the wanted apodization outside the pupil helped reduce the stroke and improve the contrast in the final DH, after stroke minimization (see Mazoyer et al, submitted in JATIS\cite{Mazoyer_JATIS}). Several solutions to relax the constrains outside the pupil are currently investigated.

\subsection{Influence of the DM setup: preliminary results}
\label{sec:comparsetup}

We also started to study the influence of the DM setup on the performance of the correction. The first important result is that the ACAD strokes are inversely related to the $D^2/Z$. Therefore, and because the stroke minimization adjustments are often negligible compared to the ACAD strokes, the final strokes are heavily dependant on this setup. The strokes indicated in Figure~\ref{fig:focalpupilplanes} (right) for the case "HiCAT bench". To obtain the contrast in the WFIRST-AFTA setup case, one have to multiply by 3 these strokes, and to divide by them by 3 to obtain the "intermediate" case.

\begin{figure}[ht]
 \begin{center}
  \includegraphics[width = 0.48\textwidth, trim= 1.3cm 0cm 1.3cm 0cm, clip = true]{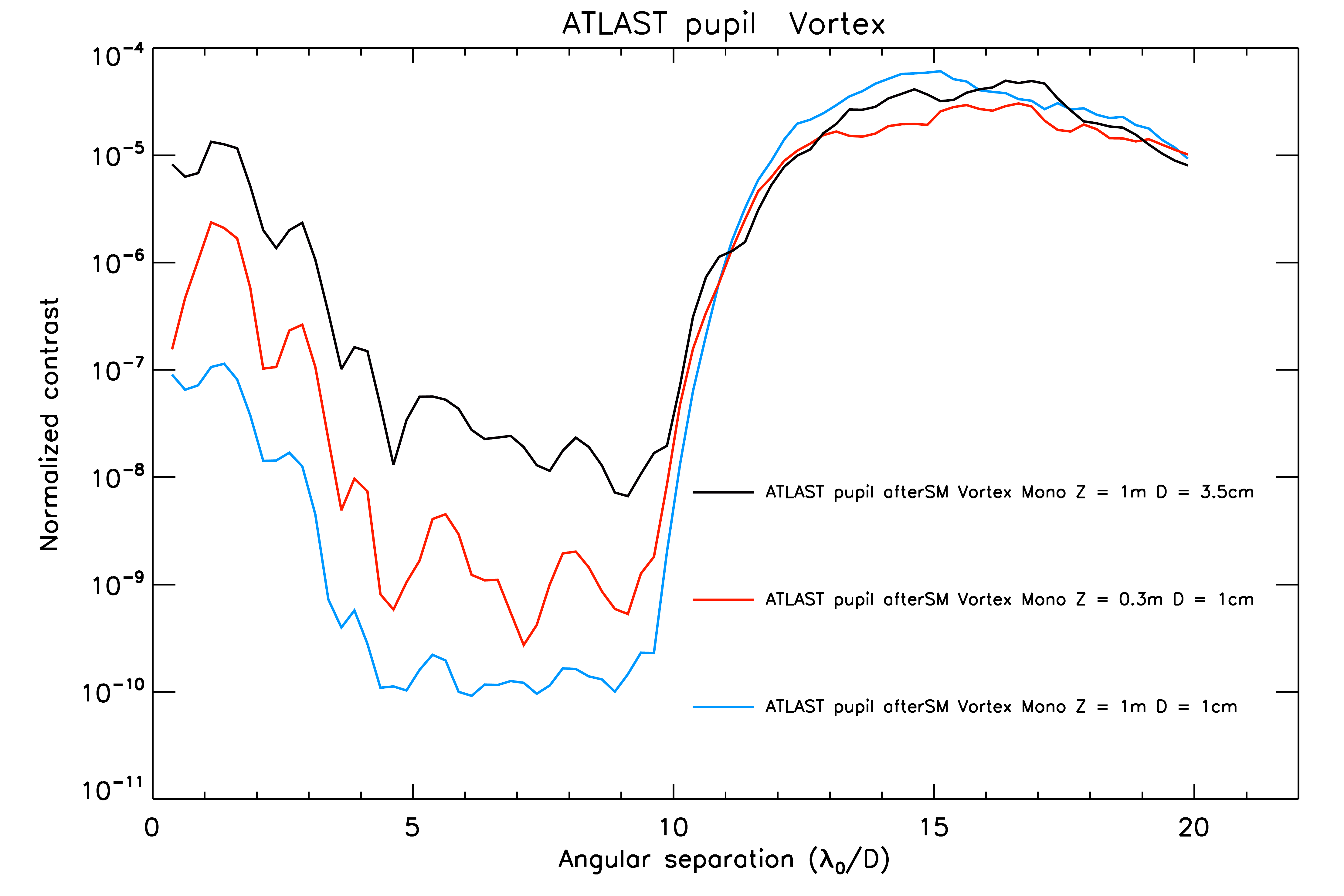}
  \includegraphics[width = 0.48\textwidth, trim= 1.3cm 0cm 1.3cm 0cm, clip = true]{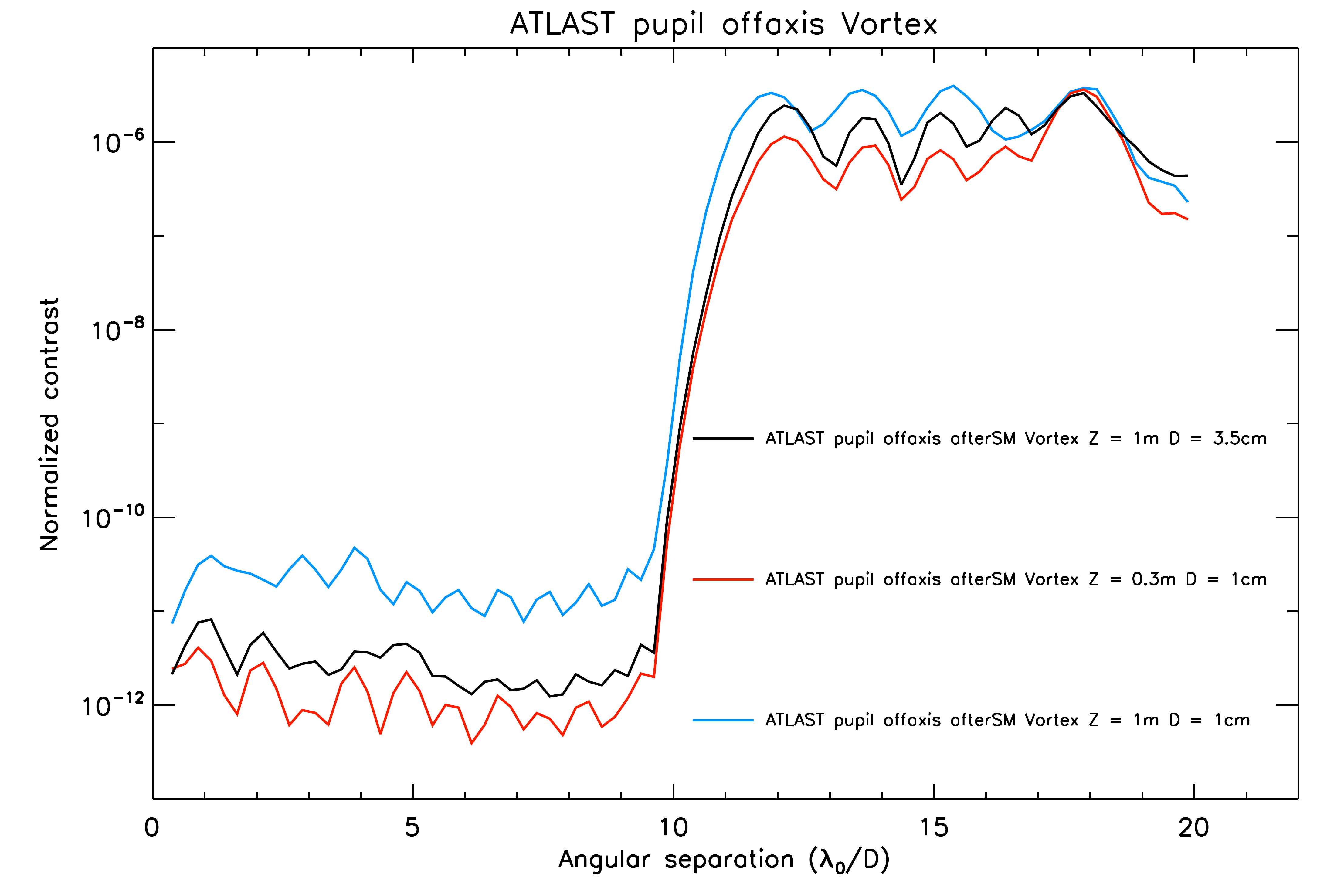}
\end{center}
 \caption[] 
{ \label{fig:contrastcurve_Vortex_comparsetup} Contrast curves for the correction with a Vortex as a function of the distance to the star in $\lambda/D$ for the ATLAST pupil on a 3 to 10 $\lambda/D$ DH (left) and the segmented off-axis pupil on a 1 to 10 $\lambda/D$ DH (right). We represented only the final contrast after ACAD + Stroke minimization in monochromatic light. In black, we plot the contrast obtained with a correction with 3.5 cm  large DMs separated by 1 m (AFTA like case). In red, we plot the contrast obtained with a correction with 1 cm  large DMs separated by 0.3 m (HiCAT like case).  In blue, we plot the contrast obtained with a correction with 1 cm  large DMs separated by 1 m (intermediate case).}
\end{figure}

Figure~\ref{fig:contrastcurve_Vortex_comparsetup} shows two cases were the setups have a different influence on the final contrast, in monochromatic correction, in the vortex case. In the ATLAST pupil case (left), the final contrast is better for high $D^2/Z$ (the best contrast level is achieved for the "intermediate" case, in blue, and the worst in the WFIRST-AFTA setup, in black). In the OAS pupil case, the "intermediate" case, in blue, is the worse.

We identified two opposing effects. The first one has been identified since Shaklan \& Green (2006)\cite{Shaklan06} who shown that for a given number of modes in the pupil the contrast is better for high $D^2/Z$. The second effect is the fact that the field on the second DM can be larger than the DM, as explained in Section~\ref{sec:method_descr}. Even if we take pupils smaller than the DMs, this phenomenon is still present and is more important if you have small $D^2/Z$. However, this is a small effect, which intervenes only for very high contrasts, as it is the case for OAS pupil. A solution in that case would be to continue to reduce the size of the pupil relatively to the DM. Simulations could help us decorrelate these two effects and find the better setups for any purpose.

\section{Conclusion}
\label{sec:Conclusion}

In this proposal, we present the preliminary results of the parametric analysis we undertook to understand completely the correction of discontinuities in the pupil with 2 DMs. These first results are promising, but we already identified several possibility of improvement.

A first short term expansion of this study is to increase the number of actuators of the DMs. Indeed, future coronagraphic instruments will use at least 48 actuator DMs (WFIRST-AFTA) of even more (64) for ELT instruments. An increase number of actuator should enhance the contrast performance in monochromatic and polychromatic light on large DH, and/or allow the DH to grow even larger. 

We did not have time to investigate the influence of jitter on the performance in our study. This parameter impact will strongly depend on the coronagraph design. The strokes (directly related to the DM setup) will also be important. We already identified solutions, introduced in Section\ref{sec:limitations} to reduce them without reducing the performance in contrast.

We will also quickly introduce other coronagraphs (fourth order vortex coronagraph, PIAA) in the simulation to take advantage of the very small IWA of these designs. We also want to understand more precisely how the DM setup influence the contrast achieve at every steps.

\acknowledgments     
 
This material is based upon work carried out under subcontract \#1496556 with the Jet Propulsion Laboratory funded by NASA and administered by the California Institute of Technology.


\bibliography{biblio_spie_sd15}   
\bibliographystyle{spiebib}   

\end{document}